\documentclass{JHEP3}
\usepackage{amsmath,amssymb}
\usepackage{graphicx}
\usepackage{array}
\usepackage{url}
\makeatletter
\def\fmslash{\@ifnextchar[{\fmsl@sh}{\fmsl@sh[0mu]}}
\def\fmsl@sh[#1]#2{
  \mathchoice
    {\@fmsl@sh\displaystyle{#1}{#2}}
    {\@fmsl@sh\textstyle{#1}{#2}}
    {\@fmsl@sh\scriptstyle{#1}{#2}}
    {\@fmsl@sh\scriptscriptstyle{#1}{#2}}}
\def\@fmsl@sh#1#2#3{\m@th\ooalign{$\hfil#1\mkern#2/\hfil$\crcr$#1#3$}}
\makeatother

\newcommand{\phan}{\phantom{3}}
\newcommand{\phant}{\phantom{3.33}}

\newcommand{\eqn}{equation}

\newcommand{\wt}{\widetilde}

\newcommand{\lb}{\left(}
\newcommand{\rb}{\right)}

\newcommand{\emiss}{\fmslash{E}}
\newcommand{\whizard}{\texttt{WHIZARD}}

\newcommand{\spheno}{\texttt{SPHENO}}
\newcommand{\sdecay}{\texttt{Sdecay}}

\newcommand{\micromegas}{\texttt{MicrOmegas}}
\DeclareMathOperator{\sgn}{sgn}

\newcommand{\GeV}{{\ensuremath\rm GeV}}

\newcommand{\ab}{{\ensuremath\rm ab}}
\newcommand{\pb}{{\ensuremath\rm pb}}

\title{Pinning down the Invisible Sneutrino}
\author{J.~Kalinowski \\ University of Warsaw, Institute of Theoretical
 Physics, PL-00681 Warsaw, Poland and Theory Division, CERN, CH-1211
 Geneva 23, Switzerland \\  
 Email: \email{jan.kalinowski@fuw.edu.pl}}
\author{W.~Kilian \\ University of Siegen, Faculty of Physics, D--57068
 Siegen, Germany \\ 
 Email: \email{kilian@hep.physik.uni-siegen.de}}
\author{J.~Reuter \\ University of Freiburg, Institute of Physics, D--79104
 Freiburg, Germany \\
 Email: \email{reuter@physik.uni-freiburg.de}}
\author{T.~Robens \\ University of Aachen, Institute of Theoretical Physics
 E, D--52056 Aachen, Germany \\
 Email: \email{robens@physik.rwth-aachen.de}}
\author{K.~Rolbiecki \\ University of Warsaw, Institute of Theoretical
 Physics, PL-00681 Warsaw, Poland \\
 Email: \email{krzysztof.rolbiecki@fuw.edu.pl}}

\abstract{ 
For points in SUSY parameter space where the sneutrino is
lighter than the lightest chargino and next-to-lightest neutralino,
its direct mass determination from sneutrino pair production process
at $e^+e^-$ collider is impossible since it decays invisibly.  In
such a scenario the sneutrino can be discovered and its mass determined from
measurements of two-body decays  of charginos produced in pairs at
the ILC. Using the event generator WHIZARD we study the prospects of
measuring sneutrino properties in a realistic ILC environment. In
our analysis we include beamstrahlung, initial state radiation, a
complete account of reducible backgrounds from SM and SUSY
processes, and a complete matrix-element calculation of the SUSY
signal which encompasses all irreducible background and interference
contributions.  We also simulate photon induced background processes
using exact matrix elements. Radiation effects and the cuts to reduce background
strongly modify the edges of the lepton energy
spectra from which the sneutrino and chargino mass are determined.
We discuss possible approaches to measure the sneutrino mass with
optimal precision.
}


\preprint{%
  CERN-PH-TH/2008-199, SI-HEP-2008-14, \\
  [0.5\baselineskip] 
  PITHA 08/23, SFB-CPP-08-68
}

\begin{document}

\section{Introduction}
\label{sec:intro}

Supersymmetry (SUSY) \cite{susy} is one of the best-motivated
extensions of the Standard Model (SM): it stabilizes the hierarchy
between the electroweak (EW) scale and the Planck scale and
naturally explains EW symmetry breaking by a radiative mechanism.
The naturalness of the scale of electroweak symmetry breaking and
the Higgs mass places a rough upper bound on the superpartner masses
of several TeV. This gives the Large Hadron
Collider (LHC) the opportunity to discover the superpartners within
the next years~\cite{lhc-tdr+susy}. Due to the LHC's hadronic environment, mostly
strongly interacting sparticles will be produced with light sleptons
and weakly interacting gauginos appearing  in decay cascades of
squarks and gluinos. Heavy weakly interacting states will be
especially hard to detect at the LHC and their
high-precision spectroscopy will only be possible at a future
high-energy International Linear Collider (ILC)~\cite{ILC}.

In this paper we focus on scenarios within the Minimal
Supersymmetric Standard Model (MSSM), in which sneutrinos are
lighter than the lightest chargino and next-to-lightest neutralino.
Sneutrino decays to charged particles in the final state are then of higher
order and therefore strongly suppressed (e.g.\ a  4-body
decay is suppressed by an off-shell intermediate  propagator and two
additional powers of the weak coupling constant). As a result,
sneutrinos decay completely invisibly into a neutrino and the
lightest supersymmetric particle (LSP, the lightest neutralino in
the considered case). A sneutrino appearing in a cascade at the LHC
is hence lost and cannot be reconstructed. An ILC threshold scan is
precluded for the very same reason. The only possibility to access
the sneutrino mass in such a case is to select a
well-reconstructible process where the sneutrino is exchanged in a
$t$-channel or shows up as a part of a cascade decay. The precise
determination of kinematic distributions gives then access to the
sneutrino mass. This idea has been proposed in
Ref.~\cite{Freitas:2005et}, where it was argued that background
effects are sufficiently under control, so a precise mass
determination is possible.

Here we study the prospects for determining sneutrino properties in
a realistic ILC environment. We include beamstrahlung, initial-state
radiation, a complete account of reducible backgrounds from SM and
SUSY processes, and a complete matrix-element calculation of the
SUSY signal which encompasses all irreducible background and
interference contributions.  Radiation, background and interference
do have considerable effect on spectrum shapes and edges, and we
discuss the possible approaches to nevertheless determine the
sneutrino mass with optimal precision.\footnote{Preliminary results
have been presented in \cite{sneutrinomassdet}.} An interesting
question is how well this can be done in a scenario with nearly
degenerate charginos and sleptons, where the corresponding decay
leptons are fairly soft and are in danger of being swamped by
photon-induced background. We restrict ourselves to areas in SUSY
parameter space where charginos are within reach of a 500 GeV ILC.

As a working point we take the SPS1a'
scenario~\cite{AguilarSaavedra:2005pw} which has been widely in use
for phenomenological studies. It is derived from the SPS1a
point~\cite{sps} by lowering the universal scalar mass parameter in
order to satisfy the cold dark matter constraint. Such a shift leads
to lower sfermion masses and, as a result, predicts sneutrinos that
decay invisibly. The SPS1a' satisfies all experimental constraints
from precision data and cosmology.

The paper is organized as follows. In Sec.~\ref{sec:process} we
discuss in detail the signal and background processes, especially
investigating whether they are experimentally distinguishable or
not. Effects coming from the inclusion of initial state radiation
(ISR) as well as beamstrahlung are discussed. The cut-based strategy
to enhance the signal-to-background ratio is developed in
Sec.~\ref{sec:cuts}, where we discuss the influence of the cut
procedure on the quality of the measurement of several observables.
The crucial observation here is the distortion of the lepton energy
spectra due to ISR and photon-induced processes as decay leptons are
quite soft in almost degenerate scenarios like SPS1a'. All analyses
in this Section have been done for a $500$ GeV center-of-mass (CM)
energy apart from Sec.~\ref{sec:800GeV}, where important differences
for an 800 GeV ILC are discussed. In Sec.~\ref{sec:mass}, we present
different approaches to the sneutrino mass determination in our
framework. Finally, we conclude as well as give an outlook.

For reader's convenience, Appendix A  recapitulates masses and
branching ratios (BRs) of supersymmetric particles for the SPS1a'
parameter point that are important for our analysis. Appendix B is
devoted to the comparison of the effective photon approximation
(EPA) with the exact matrix element method employed in our
calculations.


\section{Signal and background processes}
\label{sec:process}

After SUSY will have been found and identified at the LHC, the goal
of the ILC will be the precision determination of the masses,
quantum numbers and coupling constants in order to get access to the
high-energy theory and possibly the SUSY breaking
mechanism~\cite{AguilarSaavedra:2005pw}. Especially challenging are
measurements of sneutrino parameters in scenarios in which they
decay invisibly, as in e.g. the SPS1a' parameter point.
The difficulty lies not only in an inability of
exploiting the sneutrino pair production process, but also in
particularly many background processes, both of SUSY and SM origin,
contributing  to the signal signature as discussed below.


\subsection{Sneutrino signal and signature}\label{sec:sneutrino}

One of the standard candle processes is the pair production of the
lightest chargino,
\begin{\eqn}
\label{eq:charginopair}
  e^+e^- \,\rightarrow\, \wt{\chi}^+_1 \wt{\chi}^-_1\, \, .
\end{\eqn}
The chargino mass can be measured either by a threshold scan or in
the continuum from the edges of the decay
spectra~\cite{Freitas:2005et}. In the SPS1a' scenario the light
charginos $\wt{\chi}^+_1$ decay predominantly to $\wt{\tau}^+_1
\nu_\tau$ with a branching ratio BR($\wt{\chi}^+_1\to
\wt{\tau}^+_1 \nu_\tau$)=53.6\% (see Table \ref{tab:sps1ap} in 
Appendix A), followed by $\wt{\tau}^+_1\to \tau^+\wt{\chi}^0_1$. This
cascade produces a final state similar to that of stau pair
production. Using two tau jets in the opposite hemispheres and
missing energy as the signature for $\wt{\chi}^\pm_1\wt{\chi}^\mp_1$
production, the measurement of the chargino mass has been simulated
for the SPS1a scenario, and the expected accuracy is
$0.55$~GeV~\cite{Kato:2004bg,Weiglein:2004hn}. Detailed simulations showed that also
the stau mass in SPS1a can be measured with an uncertainty of
$0.30$~GeV~\cite{stau,Martyn}. A similar precision for the SPS1a'
point can be expected, since -- in addition -- chargino decays to
electron (or muon) and sneutrino can be exploited.

The main objective of our work, however, is to measure the sneutrino
mass, where we assume the first and second generation sneutrinos
($\wt{\nu}_e$ and $\wt{\nu}_\mu$) to be mass
degenerate.\footnote{Note, however, that the $\chi^{2}$ fit
presented in Sec.~\ref{sec:mass} can also be applied when sneutrino
masses of the first two generations are  different.} This can be
done by investigating the chargino decay modes $\wt{\chi}^+_1\to
\wt{\nu}_\ell \ell^+$, $\ell=e,\mu$,
in chargino pair production process~\cite{Freitas:2005et}. With a 
BR($\wt{\chi}^+_1\to \wt{\nu}_\ell\ell^+ $)=13\% a large sample of
events can be expected in this channel. The two-body chargino decay
leads to a uniform decay lepton energy spectrum with edges
determined by the chargino and sneutrino masses, which  provides a
method for their experimental determination.

In order to avoid large backgrounds from same-flavor lepton and
slepton production, it is advantageous to select events with leptons
of different flavor, i.e.\ one of the charginos decays to an
electron, the other to a muon. So, we search for (semi-)exclusive
final states:
\begin{\eqn}
\label{eq:mainproc}
  e^+e^- \,\rightarrow\, \wt{\chi}^+_1 \wt{\chi}^-_1\, \rightarrow
\,
  \wt{\nu}_e^*e^-\wt{\nu}_\mu \mu^+\, \rightarrow \,
  e^-\,\mu^+ \,\bar{\nu}_e \,\nu_\mu\,\wt{\chi}^0_1\,\wt{\chi}^0_1\,.
\end{\eqn}
There are also other channels that contribute to the
(semi-)exclusive final state as in Eq.~(\ref{eq:mainproc}),
\begin{\eqn}
\label{eq:irrsusy}
  e^+e^- \,\rightarrow\, X \rightarrow \,
  e^-\,\mu^+ \,\bar{\nu}_e \,\nu_\mu\,\wt{\chi}^0_1\,\wt{\chi}^0_1\, ,
\end{\eqn}
with the intermediate state $X$ that includes different production
processes as well as interference terms, e.g.~multi-peripheral
diagrams, chargino decays into neutralino and $W$, smuon
$\wt{\mu}^+_i \wt{\mu}_j^-$ and selectron $\wt{e}^+_i \wt{e}_j^-$
pair production, and single-resonant SM di-boson production. In the
following, we will distinguish between the ``proper'' signal, where
the final state products come from chargino pairs decaying to
electron and muon sneutrinos as in Eq.~(\ref{eq:mainproc}), and the
signal {\em including} irreducible background,
Eq.~(\ref{eq:irrsusy}). If not mentioned otherwise, numbers and
cross sections labeled ``signal" always include these irreducible
background processes. The mixed chargino decay modes, where one
chargino decays as $\wt{\chi}^+_1\to \wt{\nu}_\ell\ell^+ $ and the
other $\wt{\chi}^-_1\to \wt{\tau}^-_1 \bar\nu_\tau$ followed by the
cascade of $\wt{\tau}$  and leptonic $\tau$ decays, will have
additional neutrinos in the final state. Such processes are
considered as reducible background. All reducible  SUSY backgrounds
that contain additional neutrinos in the final state will be treated
in Sec.~\ref{sec:susybkgd}.

At the ILC,  for the  SUSY spectrum of SPS1a',  the Born cross
section for $\wt{\chi}_1^\pm$ pair production without initial-state
radiation or beamstrahlung is $173.56$ fb at a CM energy of 500 GeV.
The cross section reaches its maximum of $\sim$200 fb near
$\sqrt{s}=600$ GeV, and then decreases to $181.30$ fb at 800 GeV,
and falls further to $141$ fb at 1 TeV, see
Fig.~\ref{fig:chargino_xs} (tree level result shown together with
the 1-loop corrections calculated in
\cite{sneutrinomassdet,chargino_nlo,char_nlo_others}). The chargino
partial width for the decay into sneutrino and lepton is 10.2~MeV
for each of the first two generations, which constitutes 13.3\% of
the total decay width each; branching fractions for other decay
modes are 18.5\% for tau sneutrino and tau, and the dominant decay
with 53.6\% is into stau and tau-neutrino. At 500~GeV the cross
section for the proper signal $\sigma(e^+e^- \,\rightarrow\,
\wt{\chi}^+_1 \wt{\chi}^-_1) \times BR(\wt{\chi}^-_1\, \rightarrow
\, \wt{\nu}_e^*e^-) \times BR(\wt{\chi}^+_1\, \rightarrow \,
\wt{\nu}_\mu \mu^+)$ is $3.06$~fb, whereas for the six-fermion final
state, Eq.~(\ref{eq:irrsusy}), it is 4.69~fb. This shows that there
are considerable off-shell and interference effects from other
non-resonant SUSY processes contributing to the same six-fermion
final state (cf.~also~\cite{ilc_offshell}). These effects are
additionally smeared when ISR and beamstrahlung are taken into
account; we then obtain 2.50~fb for the proper signal (i.e.~when
forcing the final state to come from chargino pairs decaying to
sneutrinos) and 3.94~fb for the signal. As expected, both values are
slightly lower when ISR and beamstrahlung are taken into account, as
the emitted photons drive the  process to slightly lower values of
the effective CM energy and hence lower chargino cross section. For
a CM energy of 800~GeV, the respective numbers including initial
state radiation and beamstrahlung are 6.60~fb for the signal and
3.23~fb for the proper signal. Here, the discrepancy between full
matrix elements and narrow-width approximation is even more severe
since more SUSY processes can contribute to the exclusive final
state.

\FIGURE{
\includegraphics[width=0.6\textwidth]{sigresum}
\vspace{5mm}
\caption{\label{fig:chargino_xs} Total cross section for chargino pair
  production for the point SPS1a', at LO (medium grey, red). The NLO curves show a
  fixed order approach (black, blue) and with photon resummation (light grey, green)
  (from \cite{sneutrinomassdet,chargino_nlo}).}
}

Experimentally, the signature for the signal process is one electron
and one anti-muon and missing energy,\footnote{Equally well, one can
  consider $e^+\, \mu^-\,+\,\emiss$ as a final signature or use both
  channels.}
\begin{equation}
  \label{eq:sig}
  e^+ e^- \, \rightarrow \,e^- \mu^+ \,+\, \emiss \; .
\end{equation}
Such final states with two visible leptons and missing energy,
however, are generated by many other processes, both with
supersymmetric and Standard Model particles in the intermediate
states. Many of these processes should be suppressed both by phase
space and higher orders in the electroweak coupling constant. On the
other hand, photon-induced processes -- enhanced by large collinear
logarithms -- give by their sheer cross sections the most severe
background at the ILC. In the following, we will classify the
reducible and irreducible background processes from SM as well as
SUSY productions.


\subsection{Reducible SUSY background}
\label{sec:susybkgd}

Contamination from other SUSY processes comes from any process with
the generic signature
\begin{equation}
  e^+e^- \rightarrow \mbox{any SUSY particles} \,\rightarrow\,
  \wt{\chi}^0_1 \,\wt{\chi}^0_1
  \,e^-\,\mu^+\, + n \nu \;,
\end{equation}
with $n \geq 4$. Especially severe contaminations come from
processes that lead to two leptonically decaying $\tau$'s. They
mainly originate from the production of mixed neutralino pairs
$\wt{\chi}^0_i\wt{\chi}^0_j$,  stau pairs as well as chargino pairs.
The latter then undergo the subsequent decays to stau and LSP or tau
and sneutrino.\footnote{Since the proposed ILC vertex detector has a
  granularity that is a factor 3 finer than assumed in
  Ref.~\cite{tesla-TDR}, one could in principle use 
displaced tau decay vertices and veto against explicit tau decays. This is a
costly procedure and is to our knowledge not used in any of the
ongoing experimental ILC studies.}

In listing the contributing processes, there is a tension between
describing them signa\-ture-driven or considering them according to their
exclusive production mechanisms. Since there are big off-shell and
interference effects among several contributing weak amplitudes, as
was shown above from the cross section considerations, it is difficult
and dangerous to split processes completely into their corresponding
production channels. (Also at the ILC, SUSY processes appear to have a
generic cascade chain structure, but due to their production being
electroweak, a lot more interference among different chains is
possible and actually sizable). On the other hand, an experimental
description which takes into account only visible particles and
missing energy is not much discriminating. We here pursue the
convention to classify processes as SUSY or SM according to
whether they come from SUSY or SM production, which is apparent from
the LSP's in the final state.


\TABLE[t]
{
\begin{tabular}{|c|l|r|r|} \hline
{ Process} &
\mbox{} \hspace{1.5cm} { 500 GeV}
& $\sigma$ [fb], presel.  &  $\sigma^{\text{cut}}$ [fb]\\
\hline\hline
Signal &
$ee \to e \mu\bar\nu_e \nu_\mu
 \tilde{\chi}_1^0\tilde{\chi}_1^0$
 & $3.940(\phan8)$ & $1.639(3)$
\\ \hline\hline
SUSY $\tau$ Bkgd. &
$ee \to \tau\tau
\tilde{\chi}_1^0\tilde{\chi}_1^0 \to
e \mu
\tilde{\chi}_1^0\tilde{\chi}_1^0 4 \nu
$
 & $4.107(\phan 7)$ & $0.978(2)$
\\\hline
SUSY $\tau\nu$ Bkgd. &
$ee \to \tau\tau
\tilde{\chi}_1^0\tilde{\chi}_1^0 2\nu
\to e \mu
\tilde{\chi}_1^0\tilde{\chi}_1^0 6 \nu
$
 & $3.245(10)$ & $0.818(3)$
\\\hline
SUSY $\tau e$ Bkgd. &
$ee \to e \tau \bar\nu_e \nu_\tau
 \tilde{\chi}_1^0\tilde{\chi}_1^0\to e \mu
\tilde{\chi}_1^0\tilde{\chi}_1^0 4 \nu$
 & $3.691(\phan 9)$ & $1.102(8)$
\\\hline
SUSY $\tau \mu$ Bkgd. &
$ee \to \mu \tau \nu_\mu \bar\nu_\tau
 \tilde{\chi}_1^0\tilde{\chi}_1^0 \to e \mu
\tilde{\chi}_1^0\tilde{\chi}_1^0 4 \nu$
 & $2.617(10)$ & $0.966(8)$
\\\hline\hline
SM $WW$ Bkgd. &
$e e \to e \mu 2\nu$
 & $152.42(25)\phan$ & $0.736(2)$
\\\hline
SM $e\tau$ Bkgd. &
$e e \to e \tau 2\nu \to e \mu 4\nu $
 & $26.522(12)$ & $0.317(1)$
\\\hline
SM $\mu\tau$ Bkgd. &
$e e \to \mu \tau 2\nu \to e \mu 4\nu $
 & $15.569(54)$ & $0.174(1)$
\\\hline
SM $\nu$ Bkgd. &
$ e e \to e \mu  4\nu$
 & $0.145(\phan 1)$ &  $0.016(3)$
\\\hline
SM $\tau$ Bkgd. &
$ e e \to \tau \tau \to e \mu 4 \nu$
 & $32.679(98)$ & $<0.001$
\\\hline
SM $\tau\nu$ Bkgd. &
$ e e \to \tau \tau 2\nu \to e \mu
6 \nu$
& $3.852(10)$ & $0.335(9)$
\\\hline\hline
SM $\gamma \to \tau$ Bkgd. &
$\gamma^*\gamma^* \to \tau\tau \to e\mu 2\nu$
 & $21392(70)\phan\phan\phan$ & $0.273(2)$
\\\hline
SM $\gamma \to c$ Bkgd. &
$\gamma^*\gamma^* \to c\bar c \to e\mu jj2\nu$
 & $1089 (\phan 4)\phan\phan\phan$ & $<0.001$
\\\hline
SM $\gamma \to W$ Bkgd. &
$\gamma^*\gamma^* \to WW \to e\mu 2\nu$
 & $1.094(\phan 6)$ & $0.079(1)$
\\\hline
SM $\gamma \to \tau\nu_\tau$ Bkgd. &
$\gamma^*\gamma^* \to \tau\tau 2 \nu \to e\mu 8\nu$
 & $0.077(\phan 1)$ & $<0.001$
\\\hline
SM $\gamma\to \ell\tau$ Bkgd. &
$\gamma^*\gamma^* \to (e,\mu) \tau 2\nu \to e \mu 4\nu $
 & $0.404(\phan 2)$ & $0.055(2)$
\\\hline
\end{tabular}
\caption{Cross sections for all signal and background processes for
an ILC
  energy of 500 GeV. ISR and beamstrahlung are always included. Note
  that the final states $e$ always means electron and $\mu$ always
  anti-muon. For more details about the processes confer the
  text. The ``presel.'' column always includes a $5^\circ$ cut
  for the final electron to cut out collinear regions, and for the
  $\gamma$-induced processes a $1^\circ$ cut for particles vanishing
  in the beampipe. The last column shows cross sections after the cuts 
  discussed in Sec.~\ref{sec:cuts}. In parentheses are the {\rm
    \whizard} integration errors. 
\label{tab:500gevxsec}}
}


In the following we classify signal and background according to the
production mechanism if it is clearly identifiable (especially for
the SM processes), or specifically for the SUSY processes we use the
SM particles that appear together with the LSP in the final state
(before a final leptonic $\tau$ decay).

In our analyses we consider as (reducible) SUSY background all
processes that produce  final states with $e^-\mu^+$, two lightest
neutralinos (LSP) and four or six neutrinos. For the parameter point
SPS1a', the sum of these SUSY processes gives event numbers for the
signal signature that are bigger than the signal typically by a
factor three to four.

We distinguish the following SUSY background processes in our study,
according to the number of intermediate state $\tau$'s and
$\nu_\tau$'s, because $\tau$ decays are technically treated
differently (see below, Sec.~\ref{sec:simu}). Furthermore, the
$\tau$'s are considerably long-lived such that there is no
interference with non-resonant diagrams, and the processes can be
well split here. The largest contribution comes from two LSP's
together with a tau pair, giving rise to the two tagged leptons
$e\mu$, four neutrinos and two LSP's: $e^+e^- \to X \to \tau^+
\tau^- \wt{\chi}^0_1 \wt{\chi}^0_1 \to \wt{\chi}^0_1\,\wt{\chi}^0_1
e^- \mu^+ \nu_{\mu} \bar{\nu}_e \nu_{\tau} \bar{\nu}_{\tau}$. The
main contribution to this final state -- which we call SUSY $\tau$
background -- comes from on-shell stau pair production
$\wt{\tau}^+_i \wt{\tau}^-_j$, as well as from neutralino pair
$\wt{\chi}^0_1\wt{\chi}^0_2$ production in the intermediate state
$X$. This process receives also smaller contributions from other
intermediate $X$: neutralino pairs $\wt{\chi}^0_i\wt{\chi}^0_j$
(especially $\wt{\chi}^0_1\wt{\chi}^0_3$), production of heavy
chargino $\wt{\chi}^+_1\wt{\chi}^-_2$ (on-shell only at 800 GeV),
and even Higgstrahlung or SM di-boson production with (partial)
decay into SUSY particles.

Since we are exclusively looking into positively charged muons and
negatively charged electrons, there are processes with two different
intermediate states containing a single $\tau$: $e^+e^- \to X \to
\mu^+ \nu_\mu \tau^-  \bar\nu_\tau \wt{\chi}^0_1 \wt{\chi}^0_1  \to
\wt{\chi}^0_1\,\wt{\chi}^0_1 e^- \mu^+ \nu_{\mu} \bar{\nu}_e
\nu_{\tau} \bar{\nu}_{\tau}$, and
$e^+e^- \to X \to  e^- \bar{\nu}_e \tau^+ \nu_\tau \wt{\chi}^0_1
\wt{\chi}^0_1  \to \wt{\chi}^0_1 \wt{\chi}^0_1  e^- \mu^+ \nu_{\mu}
\bar{\nu}_e \nu_{\tau} \bar{\nu}_{\tau}$. For both of these
processes, the main contributing intermediate states are chargino
pairs $\wt{\chi}^+_1 \wt{\chi}_1^-$ with one of the charginos
decaying directly to electron or muon and the other decaying to
$\tau$. The other possible double-resonant contributions are, for
example, production of neutralino pairs $\wt{\chi}^0_1
\wt{\chi}_{3,4}^0$, stau pairs $\wt{\tau}^+_i \wt{\tau}_j^-$,  and
smuon $\wt{\mu}^+_i \wt{\mu}_j^-$ or selectron pairs $\wt{e}^+_i
\wt{e}_j^-$, respectively. Single-resonant contributions include SM
$W$ pairs, production of heavy charginos
$\wt{\chi}^+_1\wt{\chi}^-_2$, the supersymmetrized version of vector
boson fusion processes (VBF) (e.g.\ replacing the VBF final state
$e^+W^-\nu$ by $e^+\wt{\chi}^-_i\wt{\nu}$) and tri-boson production.
Because of the electron in the final state of the second process,
there are many more VBF processes and peripheral $t$-channel
topologies.

The most complex process -- the SUSY $\tau\nu$ background -- is the
one which contains in addition to the SUSY $\tau$ background also
two neutrinos in the intermediate state: $e^+e^- \to X \to \tau^+
\tau^- \nu_i \bar{\nu}_i \wt{\chi}^0_1 \wt{\chi}^0_1 \to
\wt{\chi}^0_1 \wt{\chi}^0_1 e^- \mu^+ \nu_{\mu} \bar{\nu}_e
\nu_{\tau} \bar{\nu}_{\tau} \nu_i\bar{\nu}_i$, $i = e,\mu,\tau$,
which yields 60,000 Feynman diagrams before the final $\tau$ decays.
The leading contribution is again due to production of chargino
pairs $\wt{\chi}^+_1\wt{\chi}^-_1$ and the following decays to
$\tau$ leptons. The other possibility is production of neutralino
pair $\wt{\chi}^0_1 \wt{\chi}_{2}^0$ and processes similar to SUSY
$\tau$ background with additional bremsstrahlung of electromagnetic
and weak gauge bosons. By far the highest number of diagrams is for
a heavy neutralino pair production with their subsequent decays (it
becomes important at 800~GeV, where heavy neutralinos can be
produced on-shell).

For the photon-induced SUSY processes, the largest contribution
comes from stau pairs with a cross section of 0.028~fb at a 500~GeV
ILC. Since it can be easily cut out, we do not consider any
$\gamma$-induced SUSY background from now on.


\subsection{Standard Model background processes}
\label{sec:smbkgd}

The SM backgrounds leading to the same signature of a different-flavor
opposite-sign lepton pair and missing energy (carried away by
neutrinos in the SM), mainly come from $WW$ pairs, single $W$
production and $\tau^+\tau^-$ pairs. Leptonic decays of the $W$'s and
$\tau$'s lead in all these cases to the signal signature,
Eq.~(\ref{eq:sig}). 

Pair-produced and leptonically decaying $W$'s are the most severe
background with a cross section of roughly 200 fb, which is an order
of magnitude larger than all SUSY processes. Apart from cases where
the $W$'s decay directly to an electron and an anti-muon, it includes
also processes in which one of the $W$'s or both decay to tau(s),
which then decay leptonically yielding our signature. Since the lepton
energy spectra from the $W$'s peak at half their masses, we will later
veto against very hard leptons.

\TABLE[t]
{
\begin{tabular}{|c|l|r|r|} \hline
{ Process} &
\mbox{} \hspace{1.5cm} { 800 GeV}
& $\sigma$ [fb], presel. & $\sigma^\text{cut}$ [fb]\\
\hline\hline
Signal &
 $ee \to e \mu \bar\nu_e \nu_\mu
 \tilde{\chi}_1^0\tilde{\chi}_1^0$
& $6.595(17)$  & $1.603(20)$
\\\hline\hline
SUSY $\tau$ Bkgd. &
$ee \to \tau\tau
\tilde{\chi}_1^0\tilde{\chi}_1^0
\to e\mu
\tilde{\chi}_1^0\tilde{\chi}_1^0 4 \nu
$
 & $ 3.007(\phan 6)$ & $0.731(\phan1)$
\\\hline
SUSY $\tau\nu$ Bkgd. &
 $ee \to \tau\tau
\nu_i\bar\nu_i\tilde{\chi}_1^0\tilde{\chi}_1^0
\to e\mu
\tilde{\chi}_1^0\tilde{\chi}_1^0 6 \nu
$
 & $ 4.324(16)$ & $1.147(\phan4)$
\\\hline
SUSY $\tau e$ Bkgd. &
$ee \to e\tau\bar\nu_e \nu_\tau
 \tilde{\chi}_1^0\tilde{\chi}_1^0$
 & $5.458(\phan 9)$ & $1.098(\phan6)$
\\\hline
SUSY $\tau\mu$ Bkgd. &
$ee \to \mu\tau \nu_\mu \bar\nu_\tau
 \tilde{\chi}_1^0\tilde{\chi}_1^0$
 & $3.639(\phan 9)$ & $0.974(\phan 8)$
\\\hline\hline
SM $WW$ Bkgd. &
$ee \to e\mu\bar\nu_e \nu_\mu$
 & $140.49(26)$ & $0.338(\phan 5)$
\\\hline
SM $e\tau$ Bkgd. &
$ee \to e\tau 2\nu \to e\mu 4\nu $
 & $24.330(13)$ & $0.193(\phan 1)$
\\\hline
SM $\mu\tau$ Bkgd. &
$ee \to \mu\tau 2\nu \to e\mu 4\nu $
 & $8.245(\phan 9)$ & $0.084(\phan 1)$
\\\hline
SM $\nu$ Bkgd. &
 $ ee \to e \mu \bar\nu_e \nu_\mu \nu_i \bar\nu_i$
 & $0.214(\phan 1)$ & $0.060(\phan 4)$
\\\hline
SM $\tau$ Bkgd. &
$ ee \to \tau\tau \to e\mu 4 \nu$
 &  $13.74\phan (18)$ & $0.005(\phan 0)$
\\\hline
SM $\tau\nu_\tau$ Bkgd. &
$ ee \to \tau\tau 2 \nu \to e\mu
6 \nu$
& $2.981(\phan 9)$ & $0.670(\phan 3)$
\\\hline\hline
SM $\gamma \to \tau$ Bkgd. &
$\gamma^*\gamma^* \to \tau\tau \to e^-\mu^+ 2\nu$
 & $28076 (168)\phantom{2.2}$ & $3.088(18)$
\\\hline
SM $\gamma \to c$ Bkgd. &
$\gamma^*\gamma^* \to c\bar c \to e^-\mu^+ jj2\nu$
 & $1568(\phantom{16}6)\phantom{11}\,$ & $<0.001$
\\\hline
SM $\gamma \to W$ Bkgd. &
$\gamma^*\gamma^* \to WW \to e\mu2\nu$
 & $3.458(21)$ & $0.620(\phan 4)$
\\\hline
SM $\gamma \to \tau\nu_\tau$ Bkgd. &
$\gamma^*\gamma^* \to \tau\tau 2 \nu \to e\mu 8\nu$
 & $0.116(\phan 1)$ & $0.0410(\phan 2)$
\\\hline
SM $\gamma\to \ell\tau$ Bkgd. &
$\gamma^*\gamma^* \to e^- \tau^+ 2\nu \to e^- \mu^+ 4\nu $
 & $1.256(\phan 6)$ & $0.356(\phan 1)$
\\\hline
\end{tabular}
\caption{The same as in Table~\ref{tab:500gevxsec} but for an ILC
  energy of 800 GeV.
\label{tab:800gevxsec}}
}

As a second severe SM background, we have leptonically decaying
$\tau$ pairs. Their cross section is 32.7~fb, being an order of
magnitude larger than the SUSY signal, and still a factor three
larger than all SUSY processes. Since the leptons originating from
tau decays have a genuine back-to-back structure, this enables one
to identify the tau background with a thrust-like variable.

At the ILC there are also photon-induced SM processes that
constitute a severe background \cite{Martyn,tesla-TDR,Zhang:2007nz}.
This background is usually two to four orders of magnitude larger
than the corresponding signal one is interested in. The background
stems from collinear photon radiation off the incoming electrons and
positrons, which do not get a large $p_\perp$ kick and vanish in the
beampipe. The collinear photons then trigger very much the same SM
background processes, but compared to the genuine electroproduction
the cross sections -- although reduced by the square of the
electromagnetic coupling constant -- are enhanced by large collinear
logarithms.

Although the designated electromagnetic calorimeters at the ILC will
have a spectacular resolution in the extreme forward and backward
directions, we demand the detected electron in all cases to have a
polar angle $\theta_e\,>5^\circ$ with respect to the beam axis. This
cuts out the collinear singular region, where perturbation theory is
no longer reliable. For the photon induced processes, particles with
a polar angle $<\,1^{\circ}$ with respect to the beam direction will
be treated as lost in the beampipe. Therefore, we preselect events
by requiring $\theta_e\,>\,5^{\circ}$, while for other charged
particles considered to be lost in the beampipe -- jets and
additional leptons -- we demand $\theta\,<\,1^{\circ}$.

While the total cross section for photon induced processes is huge
(about 12~nb in total, about 900~pb for charm, 300~pb for tau, and
90~fb for $W$ pair production), it greatly reduces with the decay
branching fractions to the required final state folded in, and after
preselection. Nevertheless, these processes are still very large,
amounting to roughly 23~pb at a 500~GeV ILC. By far the most
dominating background is from photon-induced tau pairs decaying to
the tagged leptons with a cross section of 21.4~pb. Their broad
spectrum at low energies together with soft leptonic decays distorts
the shapes of the low-energy edges of the electron and muon
distributions that are important for the chargino and sneutrino mass
determination.

Photon-induced charm pairs have even larger total cross section than
the $\tau$'s, but vetoing against jet activity in the central
detector from their semi-leptonic decays cuts this down to a value
of 1.1~pb, which is still a factor 250 larger than the signal.
Leptonically decaying $W$s contribute another 1.1~fb.

The cross sections for the signal, the SUSY and SM backgrounds after
preselection for $\sqrt{s}=500$~GeV ILC are collected in
Table~\ref{tab:500gevxsec}. The right-most column in
Table~\ref{tab:500gevxsec} shows the change in cross sections after
applying suitably chosen cuts which will be discussed in detail in
Sec.~\ref{sec:cuts}. The corresponding values for $\sqrt{s} =
800$~GeV are given in Table~\ref{tab:800gevxsec}.


\subsection{Event simulation}
\label{sec:simu}

For the simulation of the SUSY signal and SUSY/SM background
processes, we take the multi-purpose event generator
\whizard\ \cite{whizard}, which is especially suited for beyond
the SM applications and well-established for SUSY
simulations~\cite{whizard_bsm}. It allows for the usage of full matrix elements for exclusive
final state particles, and automatically generates all contributing
intermediate (on- and off-shell) states. Hence, intermediate
resonances as well as interferences from off-shell continua are
considered on equal footing. For ILC, the inclusion of off-shell
states in full matrix elements is mandatory, especially when cuts have
to be taken into account~\cite{ilc_offshell}.

Since the process considered here (as most electroweak SUSY
processes and SUSY decay cascades) are dominated by tau leptons, it
is crucial to simulate leptonic decays of taus. Since taus are very
narrow resonances, a multi-channel adaptive phase space routine has
difficulties finding the corresponding poles. In that case, a
narrow-width approximation for the tau decays is
appropriate~\cite{tauola}. We simulated the energy and angular
distributions of the leptonically decaying $\tau$'s and the
semi-leptonically decaying charm quarks by extending the \whizard\
program accordingly.

As mentioned above, the most severe background at a high-energy lepton
collider comes from photon-induced processes. Such contributions are
commonly accounted for by using the equivalent photon approximation
(EPA)~\cite{EPA,Budnev:1974de}, which approximates the collinear
radiation by a structure function approach of on-shell photons inside
the electrons. However, the EPA tweaks the
kinematics and results in deviations in total cross sections
as well as in the shapes of differential distributions.
This is particularly important when kinematic cuts are used to
select the desired signal. Taking the photon-induced tau-pair production process as an example,
in Appendix B we compare EPA with the exact
treatment exposing
the difference in the lepton energy spectra before and after cuts.
In \whizard, the approximation of on-shell intermediate photons is
avoided by using exact matrix elements for collinearly radiating
electrons that vanish in the beam pipe. However, it is necessary to
run \whizard\ in a higher (quadruple) floating point precision for
taking properly into account the large $p_\perp$ logarithms of the
radiating electrons.

Since all distributions at the ILC are heavily dependent on photon
initial state radiation as well as beamstrahlung effects, it is
absolutely mandatory to include these effects. This is done using
the setup inside the \whizard\ generator, which employs an ISR
structure function that resums all leading soft- and soft-collinear
logarithms and takes into account hard-collinear terms up to the
third order~\cite{Jadach}. The beamstrahlung is simulated online
using the \texttt{CIRCE} package~\cite{circe} that comes with the
\whizard\ distribution. In~\cite{ilc_offshell} it has been shown
that ISR and beamstrahlung can heavily distort the box-shaped
spectra of energies from decay products of heavy states whose mass
one wants to determine. Since at 500 GeV the collider energy is at
the rising shoulder of the production cross section for charginos in
the SPS1a' scenario, the radiative return slightly reduces the
number of signal events, bringing the signal process closer back to
the threshold. On the other hand, the SM background, dominated by
the $WW$ production, is slightly enhanced by the radiation effects.
Therefore in the following, beamstrahlung and ISR are always taken
into account.

Although we did not study effects of final state radiation (FSR), we
briefly want to comment on its possible effects. From fast
quasi-massless fermions like electrons and muons, final state
collinear photon radiation could be enhanced by  $\log\lb E_\ell /
\delta_\ell\rb$, where $\delta_\ell$ is the energy resolution for
collinear photons. Since for the near-degenerate SPS1a' spectrum the
decay leptons are rather low-energetic, we do not expect this to be
a sizable effect.


\section{Cut-based strategy for enhancing the signal to background ratio}
\label{sec:cuts}

Since the final state $e^- \mu^+ \,+\, \emiss$ contains only two
visible particles, it might seem to provide only a limited amount of
experimental information. Nevertheless, several experimental
observables can be exploited to enhance the signal to background
(S/B) ratio by applying suitable cuts on the lepton polar angle
$\theta(\ell)$ (measured from the $e^-$ beam direction), on energy
$E(\ell)$ and transverse momentum $p_\perp(\ell)$ distributions, on
azimuthal angular separation of outgoing leptons
$\Delta\phi(e,\mu)$, as well as on missing energy and momentum.


\subsection{Cutting out the SM and SUSY background}
\label{sec:cutsec}

The cuts to enhance the S/B ratio have been optimized for a CM
energy of 500~GeV, with slight modification for 800~GeV
(cf.~Sec.~\ref{sec:800GeV}).

As was discussed in Sec.\ \ref{sec:smbkgd}, after preselection the
most severe background comes from photon-induced SM processes which
are larger than the signal 
by a factor $10^4$. Enhanced by large collinear logarithms, these
processes contribute to the final state leptons from the leptonic
$\tau$ and semi-leptonic charm decays. Hence, $e$ and $\mu$ appear
soft and along a thrust-like axis parallel to the beam with a large
azimuthal separation. Therefore, a cut on the total transverse
momentum $|\vec{p}_\perp (e) + \vec{p}_\perp (\mu)| > 4$ GeV (and
also on the separate $p_\perp(e,\mu)>2$ GeV) together with a cut on
the azimuthal separation of the leptons $|\Delta
\phi(e,\mu)|<150^\circ$ provide a very powerful means to reduce the
gamma-induced background by roughly two to three orders of
magnitude. [Note, that a different strategy was followed by the LEP
experiments, namely to determine the gamma-induced background from a
control sample by tagging the emission of an additional hard
photon~\cite{lepsusywg}.] In addition, the azimuthal correlation cut
on the $e, \mu$ system eliminates the SM $\tau$ background, since
the $\tau$'s are heavily boosted.

Background from $W$-pair production contributes to the higher-energy
parts of lepton spectra, since the decay leptons shape a Jacobian
peak at approximately half of the $W$ mass. The considerable boost
of the $W$'s shifts this peak further up. This argument applies to
both direct leptons as well as to $e, \mu$ coming from $W$'s going
into leptonically decaying $\tau$'s. Therefore, constraining final
state leptons to the energy window 1 $<E(e,\mu)<$ 40 GeV reduces the
$W$ backgrounds (which is the largest SM background after cuts), by
two orders of magnitude.

Furthermore, we require two well-visible leptons in the central part
of the detector, i.e.\ $15^\circ < \theta(e^{-}) < 155^\circ$ and
$25^\circ < \theta(\mu^{+}) < 165^\circ$. The complete set of cuts
reduces the SM backgrounds by four orders of magnitude, yielding a
S/B ratio of order 1. All cuts are summarized in
Tab.~\ref{tab:cutsall}, and their effects on signal and background
processes are collected in Tab.~\ref{tab:cuteffs}. In the latter
table cross sections are shown in units of fb. In the second
column cross sections after preselection are shown, in the following
columns when relaxing the corresponding cut(s). The most efficient
cuts for the individual processes are indicated by shading the
corresponding entry in grey. The last column shows the result after
applying all cuts. Note that a cut on missing energy or missing
transverse momentum does not help in enhancing further the SUSY to
SM ratio since there is severe background from $Z\to \nu\bar\nu$ and
vector boson fusion.

\TABLE[t]
{
\begin{tabular}{|l|cc|cc|}
\hline $\sqrt{s}$, \hspace{1cm} Cut
& 500, lower & 500, upper & 800, lower & 800, upper \\
\hline\hline
  $E(e^-)$, $E(\mu^+)$ & 1 GeV & 40 GeV & 1 GeV & 60 GeV
\\
  $\theta(e^-)$   &   $15\,^\circ$   &   $155\,^\circ$
&  $25\,^\circ$   &   $165\,^\circ $
\\
  $\theta(\mu^+)$ &   $25\,^\circ$   &   $165\,^\circ$
&  $10\,^\circ$  &   $165\,^\circ$
\\
  $\Delta\phi(e^-,\mu^+)$   &   $-150\,^\circ$   &   $150\,^\circ$
&  $-150\,^\circ$  &   $150\,^\circ$
\\
  $p_\perp(e^-)$, $p_\perp(\mu^+)$  & 2 GeV & & 2 GeV &
\\
  $|\vec{p}_\perp(e^-) + \vec{p}_\perp(\mu^+)|$ & 4 GeV &  & 4 GeV &
\\\hline
\end{tabular}
\caption{\label{tab:cutsall} Cuts used in the analysis for an ILC
with
  500 GeV and 800 GeV CM energy, respectively.}
}

\TABLE[t]
{
\includegraphics[width=.98\textwidth]{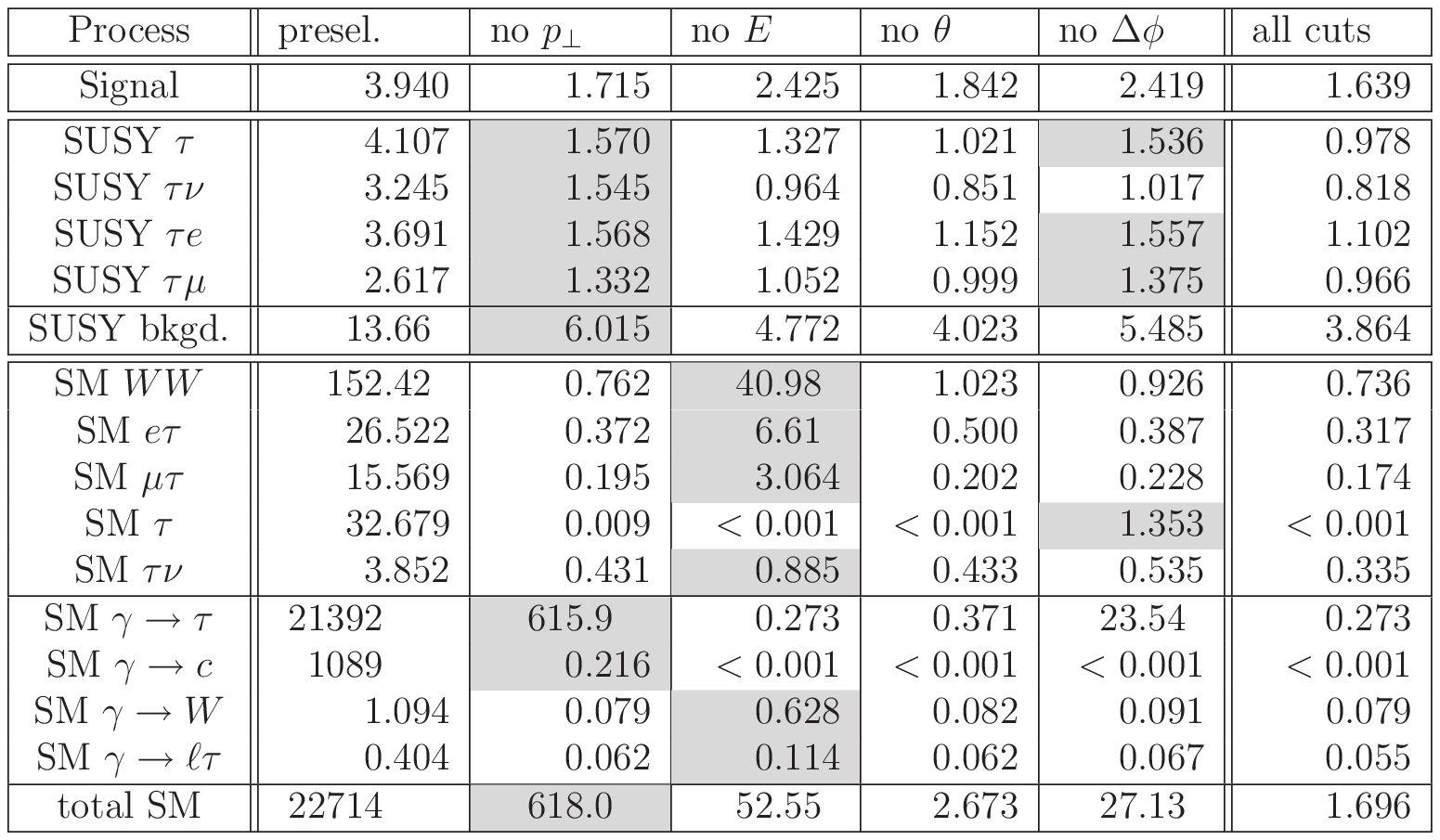}
\vspace{-5mm}
\caption{\label{tab:cuteffs} Effects of cuts on the signal and
background processes at CM energy 500~GeV. All values are given in fb.
Description of columns is given in the text; the
last column corresponds to the cuts of Tab.~\ref{tab:cutsall}. The
most important cuts for the individual processes are shaded in
grey.}
}

Next, we investigate how the cuts affect the SUSY processes. There
are two issues here: one is the separation of the signal process
from reducible SUSY backgrounds, the other is a desired enhancement
of the rate of proper signal over irreducible SUSY backgrounds. The
latter is important, since we want to apply kinematic analysis to
determine sneutrino and chargino parameters, which mostly rely on
the on-shell kinematics of the proper signal process
(cf.~Sec.~\ref{sec:mass}).

SUSY cascade decays for parameter points with an almost degenerate
spectrum of weakly interacting states, like SPS1a', generate
relatively soft leptons with an energy of a few up to few tens of
GeV. Since the leptons here originate from decays of quite heavy
states, they are radially distributed in the detector, and electron
and muon are almost angular uncorrelated. Due to the complexity of
electroweak SUSY amplitudes and their corresponding phase space
configurations, there is in general no single well-motivated cut for
reducible SUSY background. Interfering cascades, especially
semi-resonant processes, with only small differences in kinematics
are difficult to cut out. Nevertheless, with the set of our cuts
these backgrounds are reduced by a factor of three to four.

The irreducible SUSY background is also reduced, although to a
lesser extent: a 40\,\% contamination from non-resonant or partly
resonant contributions is brought down to a mere 13\,\%. This is
more pronounced for the muon distribution
(Fig.~\ref{fig:propervsfull}) than for electrons in the energy range
left over after the cuts. Most of the irreducible background, which
comes from multi-peripheral $t$-channel diagrams, is cut out by the
upper energy cut of the $e^-$. The rate of the proper signal itself
suffers mostly from the $\Delta \phi$, but only moderately.
Nevertheless, after all cuts we have a clearly visible SUSY signal
with S/B $\sim$ 3/1.

\FIGURE{
    \includegraphics[width=.42\textwidth]{1_500_nc.1}\qquad\qquad
  \includegraphics[width=.42\textwidth]{1_500_ac+.1}\\[10mm]
    \includegraphics[width=.42\textwidth]{1_500_nc.2}\qquad\qquad
    \includegraphics[width=.42\textwidth]{1_500_ac+.2}
    \vspace{2mm}
\caption{\label{fig:propervsfull} Electron (upper row) and muon
(lower row) energy distributions for the proper signal (dark grey,
blue) and full signal with irreducible SUSY background (dark+medium
grey, blue+green). Left panels show histograms after preselection,
and right panels after all cuts. In the muon distribution, the ratio of proper over full signal is
considerably enhanced by the cuts.}
}


\subsection{Cut effects on lepton energy distributions}
\label{sec:cutedge}

For determination of sneutrino and chargino masses from kinematic
variables at the ILC, the energy distributions of the decay leptons
are crucial. It is important to realize that these are distorted by
ISR, beamstrahlung and selection cuts, as well as by
contributions from background events that survive cuts. The aim of
this section is to discuss the consequences of these distortions. When
plotting only the proper 
signal (i.e.~restricting to intermediate charginos decaying to 
sneutrinos), the famous on-shell box shape of the lepton spectrum is
clearly visible, Fig.~\ref{fig:propervsfull}.\footnote{In all
  histograms the number of events per 1 GeV bin assuming an integrated
  luminosity of 1 ab$^{-1}$ is shown.} The cuts applied to 
enhance the S/B ratio do not do much harm to this feature: although
smeared to a certain degree (more for electrons than for muons) the
box character is still visible.


\FIGURE[ht]{
    \includegraphics[width=.42\textwidth]{energiesetc.1}
    \qquad\qquad
    \includegraphics[width=.42\textwidth]{energiesetc.2}
    \vspace{2mm}
\caption{\label{fig:lep500} Lepton energy distributions after
applying the cuts of Tab.~\ref{tab:cutsall}: SM background (light
grey), SUSY background (medium grey, red) signal (dark grey, blue)
for the  electrons (left panel) and muons (right panel).}
}

\FIGURE[t]{
  \begin{tabular}{cc}
    \includegraphics[width=.36\textwidth]{1_23_nc.1} & \qquad\quad  
    \includegraphics[width=.36\textwidth]{1_23_ac+.1} \\ 
    & \\ \vspace{2mm}
    \includegraphics[width=.36\textwidth]{1_23_nc.2} & \qquad\quad
    \includegraphics[width=.36\textwidth]{1_23_noptnophi.2} \\
    & \\ \vspace{2mm}
    \includegraphics[width=.36\textwidth]{1_23_nophic.2} & \qquad\quad
    \includegraphics[width=.36\textwidth]{1_23_ac+.2}    
  \end{tabular}

\caption{\label{fig:cut_lepdistr} Lepton energy distributions for
the signal (dark grey, blue) and reducible SUSY $\tau e$ background
(medium grey, red). Upper row: electrons after preselection (left
panel) and with all cuts (right panel). Remaining plots are for
muons: after preselection (middle, left), after energy and angle
cuts (middle, right), after energy, angle and $p_\perp$ cuts (lower,
left), and after all cuts (lower, right). The low-energy edge of the
box-type energy distribution of muons is no longer visible after the
$p_\perp$ cut. Note different vertical scales in these plots.}
}

The lepton spectrum from the SM background after cuts is rather
flat, the light grey histograms in Fig.~\ref{fig:lep500}, which
allows one to subtract it in the analysis. The SUSY signal sits
nicely on top of it, making a distinction between SUSY and SM quite
easy. On the other hand, it is not possible to discriminate between
SUSY background (reducible and irreducible) and the proper signal
with any kind of cuts. Especially the similarities in the kinematic
set-up of SUSY $\tau$ and SUSY $\tau\ell + \tau\nu$ backgrounds make
it difficult.

To expose our point better, let us consider a specific SUSY $\tau e$
background process
\begin{equation}
\label{eq:proc23}
   e^{+} e^{-} \longrightarrow \wt{\chi}^{+}_{1} \wt{\chi}^{-}_{1}
   \longrightarrow e^{-}\wt{\nu}^*_{e} \tau^{+}
   \wt{\nu}_{\tau} \longrightarrow e^{-} \bar{\nu}_{e} \mu^{+}
   \nu_{\mu} \nu_{\tau} \bar{\nu}_{\tau} \wt{\chi}^{0}_{1}
   \wt{\chi}^{0}_{1}\; ,
\end{equation}
where one of the charginos decays directly into $e$ and $\wt{\nu}_e$,
while the final muon is the decay product of the tau
coming from the other chargino. The same discussion applies to the case with
$e,\mu$ flavors exchanged, and for the other $\tau$-induced
SM or SUSY background processes.

Figure \ref{fig:cut_lepdistr} shows the lepton spectra for the full
signal, Eq.~(\ref{eq:irrsusy}) (dark grey, blue), and for the SUSY
$\tau e$ background, Eq.~(\ref{eq:proc23}), (medium grey, red); in
these plots SM background has been removed. For the electron
spectrum the SUSY $\tau e$ background does not impair the edges. On
the contrary, it enhances the number of signal-like events, see the
first line of Fig.~\ref{fig:cut_lepdistr}. This should not be
astonishing since the SUSY $\tau e$ (and $\tau \mu$) background
processes  share in one hemisphere the same chargino cascade decay
chain with the signal process. As a result, for electrons the
kinematics in both cases is almost identical.

For the muons in this particular process the situation is
quite different. Here the spectrum at the lower edge of the box
eats into the raising infrared tail of the SUSY $\tau e$ background.
The middle and lower lines of Fig.~\ref{fig:cut_lepdistr} show the
successive application of the cuts listed in the previous section to
eliminate the SM backgrounds. After the $p_\perp$ and $\Delta\phi$
cuts, the lower edge of the muon spectrum is extinct; releasing the
$p_{\perp}$ cut, however, reintroduces a large SM background
(cf.~Sec.~\ref{sec:cutsec}).

The complete lepton spectra with signal, remaining SUSY background
as well as SM backgrounds are shown in Fig.~\ref{fig:lep500}. The
lower edge, which in the signal is due to the sneutrino decay, is
essentially smeared out. The upper edge is more pronounced. We will
discuss how to determine the sneutrino mass with this complication
in Section~\ref{sec:masscross}.


\subsection{Increasing the collider energy}\label{sec:800GeV}

It is interesting to investigate  possible benefits of increasing
the collider CM energy to 800 GeV.  In this case, the SPS1a'
scenario exposes  different phenomenological features than at lower
CM energy. On the one hand, the chargino pair production is bigger
by a factor of about 1.5 with respect to 500~GeV, see
Fig.~\ref{fig:chargino_xs}, which increases the signal statistics. At the
same time, however, the irreducible background is also bigger, since
now many new SUSY intermediate states
are either above or close to their thresholds. The same argument
applies to most of the reducible SUSY background, while the
$\tilde{\tau}$ pair production is slightly smaller. The SM
background is also reduced: the $WW$ cross section falls by 10\%,
$\tau\mu$ by a factor 2, while the $\tau$ pair production cross
section is smaller by a factor of three. On the other hand, we are
now on the falling shoulder of the chargino pair production cross
section, so we experience larger effects of beamstrahlung and ISR
similar to a radiative return effect. The radiative return also
enhances the photon-induced $\tau\tau$ by 30\% and $WW$ roughly by a
factor 3.

The result for all SUSY signal and background processes for a center
of mass energy of $800\,\GeV$ before and after cuts can be found in
Table~\ref{tab:800gevxsec}. Note that the cuts here are adapted to
the higher energy: the upper edge of the lepton energies was raised
to 60 GeV, since the upper edge at this energy is at $E\sim 45$ GeV.
Also less restrictive polar angle cuts for muons
have been applied, see
Table~\ref{tab:cutsall} for details. Comparison of last columns of
Tables \ref{tab:500gevxsec} and \ref{tab:800gevxsec} shows that
increasing the ILC energy does not change  cross sections after cuts
by a large factor. They are of similar order.

\FIGURE{
    \includegraphics[width=.42\textwidth]{energiesetc800.1}
    \qquad\qquad
    \includegraphics[width=.42\textwidth]{energiesetc800.2}
    \vspace{2mm}
\caption{\label{fig:lep800} Lepton energy distributions
for electrons (left panel) and muons (right panel) after applying
the cuts of Tab.~\ref{tab:cutsall} for $\sqrt{s}\,=\,800\,\GeV$; SM
background (light grey), SUSY background (medium grey, red) and the
signal (dark grey, blue). }
}

\FIGURE{
    \includegraphics[width=.42\textwidth]{1_800_nc.2}\qquad\qquad
    \includegraphics[width=.42\textwidth]{1_800_ac+.2}
    \vspace{2mm}
\caption{\label{fig:emu_1_800_red} Muon energy distribution for the
  proper signal (dark grey, blue) and irreducible background (medium
  grey, green) at $\sqrt{s}=800$~GeV; left: without cuts, right: with
  cuts. It is clearly visible that the ratio of the irreducible
  background to proper signal is greatly reduced.} 
}

More dramatic change, however, is observed in the lepton energy
distributions. With increased collider energy the infrared and
collinear logarithms are much more pronounced. As a result, the
low-energy tails are basically parallel for SM and SUSY, which means
especially that the SM background from photon-induced $\tau$'s is no
longer flat. As can be seen in Fig.~\ref{fig:lep800}, after the cuts
from Tab.~\ref{tab:cutsall} all spectra look quite similar.
Although after cuts the background is reduced by a factor $10^{4}$ and
the S/B ratio is still quite good $\sim 1$, the shapes of 
the distributions are much less favorable. The exponentially
rising low energy structure from leptonic $\tau$ decays overlap with
the crucial lower edge region. Moreover, not only the lower, but
also the upper edge is no longer clearly detectable.

To understand the origin of stronger smearing effect for the upper
edge of energy distribution, let us assume that the SM background
can be measured well enough (presumably from continuum measurements
below/far away from SUSY thresholds) and subtracted. The muon energy
distributions for SUSY processes only are shown in
Fig.~\ref{fig:emu_1_800_red}. Before cuts  the box-shape character
is still visible although  the upper edge is less pronounced than at
500~GeV. Roughly half of the signal consists of irreducible
background, which  goes down to about 25\% after applying cuts to
suppress the SM background. However, at the same time the upper edge
is killed. This is mainly due to the $\Delta\phi$ cut. With
increased collider energy charginos are more boosted producing more
events with  back-to-back leptons. Such events are removed by the
$\Delta\phi$ cut, which however cannot be relaxed because of the
necessary suppression of the huge background from photon induced
$\tau$ pair production. A similar effect, however not that strong, is
visible in Fig.~\ref{fig:cut_lepdistr} at CM energy 500 GeV. 

The above discussion shows the importance of adjusting the collider
energy to the spectrum. Increasing the ILC energy is more desirable to
detailed studies of heavier SUSY particles.


\section{Sneutrino mass determination}
\label{sec:mass}

\subsection{Sneutrino mass determination from total cross sections}
\label{sec:masscross}

In the chargino production and decay process $e^+e^-\to
\tilde{\chi}^+_1\tilde{\chi}^-_1\to e^-\mu^+ + \emiss$ the
dependence on the sneutrino mass enters both in the production and
decay matrix elements. In the production process the electron
sneutrino is exchanged in $t$-channel\footnote{ This dependence
can also be exploited using forward-backward asymmetries, see
\cite{Desch:2006xp} for details.}. For SPS1a' point this leads
to moderate dependence of the cross section on this parameter, which
amounts to a $1\%$ change of the cross section for a 1 GeV change in
the sneutrino mass at $\sqrt{s}=500$~GeV. At the same time the
sneutrino mass enters the decay matrix element, since the two-body
decay channel $\wt{\chi}_1^+ \to \wt{\nu}_\ell \ell^+$ is open. For
the SPS1a' scenario approximately $50\%$ of charginos decay to one
of the sneutrinos (cf.\ Appendix \ref{sec:sps1ap}).

Specifically, in the current analysis we concentrate on chargino
decays to electron and muon sneutrinos. We find  that for this
particular final state, Eq.~(\ref{eq:mainproc}), the sensitivity of
the production cross section (together with irreducible background)
on the sneutrino mass amounts to $\sim 25\%/$GeV, which is much
stronger comparing to the sole production process of charginos. This
is the consequence of strong kinematic effects in the decay
amplitude when the parent and daugther particles, chargino and sneutrino, are close in
mass. Therefore looking at this particular channel gives a very
good opportunity to determine the sneutrino mass.

Since experimentally we cannot distinguish chargino production and
decay process from the reducible and irreducible SUSY background,
the latter also have to be included in the analysis. This will
dilute the strong effect coming from the signal. However, since some
of these background processes (e.g.\ SUSY $\tau e$, $\tau \mu$ and
$\tau\nu$) also involve chargino decays mediated by on-shell
sneutrinos, we can still expect quite strong dependence. Another
major source of uncertainty that has to be taken into account in
determining the value of the  cross section, apart from the
sneutrino mass, comes from the error on chargino mass. Assuming that
the chargino mass can be measured elsewhere with a precision of
1~GeV, it gives an uncertainty of signal and SUSY background cross section after cuts
of order $0.7$~fb (i.e.\ $\sim 17\%/$GeV). This high
sensitivity can again be attributed to the kinematic effects appearing in
the chargino decay matrix element.

\FIGURE{
    \includegraphics[width=0.42\textwidth]{signalcut} \qquad
    \includegraphics[width=0.42\textwidth]{susycut}
    \vspace{2mm}
\caption{\label{fig:signalnum} Total number of events after cuts as
a function of sneutrino mass for signal only (left) and with SUSY
backgrounds included (right). The horizontal lines give 1-sigma
statistical error for the nominal SPS1a' sneutrino mass (172.52~GeV)
and for the integrated luminosity $\mathcal{L}=1\,\ab^{-1}$.}
}

A fit of the collider data to different samples of
Monte Carlo data for varying (electron and muon) sneutrino masses
can thus be used to get an estimate of the sneutrino mass.
Figure~\ref{fig:signalnum} shows the number of events after
all cuts for the signal (left) and  for signal and SUSY background
(right). We assume that the SM background, which is flat, is
understood well enough  and can be subtracted. 
Our ``collider data'' are the MC sample for the nominal SPS1a' mass.
The characteristic dependence on the sneutrino mass is apparent,
although the curve becomes flatter when including the SUSY
background. The SUSY background has been generated with a varying
sneutrino mass as well. Taking into account statistical error and the
uncertainty due to the  chargino
mass error, this simple method allows to determine the sneutrino
mass range of roughly
\begin{equation}
m_{\tilde{\nu}} \sim 172.50 \pm 0.75 \mbox{ GeV}\;.
\end{equation}
No other systematic or parametric errors have been included here. In
principle, generation of the MC samples with varying sneutrino
masses demands the knowledge of considerable parts of the SUSY
parameters. This inverse problem can be solved by fitting the
experimental data in global fits to the corresponding
models~\cite{Rauch:2007xp,Bechtle:2005qj}. Nevertheless, our
simplified method already allows us  to narrow down the possible
region of the sneutrino mass. A more spectrum-/model-independent
way of sneutrino mass determination using kinematic
methods is discussed in the next subsections.


\subsection{Sneutrino  mass determination from
  on-shell kinematics}

Pure kinematic relations can be exploited to
determine the masses of sneutrino and, eventually, chargino in a
model-independent way. Masses of these two 
superpartners can be determined from kinematics of on-shell chargino
production followed by the on-shell 2-body $\wt{\chi}^\pm_1\to
\wt{\nu}_\ell\ell$ decay~\cite{Freitas:2005et,Datta}. For the given
incoming $e^+e^-$ energy the lepton decay spectra are uniform
between the two energies $E_{\rm min}$ and $E_{\rm max}$, given by:
\begin{equation}
E_{\rm min, max}= E^*_\ell \frac{1\pm \beta}{\sqrt{1-\beta^2}}\;,
\end{equation}
where $E^*_\ell = (m_{\wt{\chi}}^2-m_{\wt{\nu}}^2) / (2
m_{\wt{\chi}})$ is the lepton energy in the chargino rest frame and
$\beta=(1-4m^2_{\wt{\chi}}/s)^{1/2}$ is the chargino velocity
in the CM system. Inverting, we get
\begin{equation}
  \label{eq:ayresrel}
  m_{\wt{\chi}^{\pm}_1} = \sqrt{s} \frac{\sqrt{E_\text{min}
      E_\text{max}}}{E_\text{min} + E_\text{max}}\;,
  \qquad
  m_{\wt{\nu}_\ell} = m_{\wt{\chi}^{\pm}_1}
  \sqrt{1-\tfrac{2 (E_\text{min} + E_\text{max})}{\sqrt{s}}}\;.
\end{equation}

The chargino polarization
effects~\cite{Freitas:2005et,Datta,Choi:2000ta}, which are all
included in our calculations, affect the uniform distributions and
their response to kinematical cuts. Beam- and bremsstrahlung also
modify the shape of the energy distributions. Other sources of
smearing come from off-shell and interference effects, which however
are not overwhelmingly important since for the SPS1a' point
charginos and sneutrinos are very narrow. Finally, irreducible SUSY
and SM backgrounds, as discussed in the previous section, make the
determination of the edges of lepton energy spectra more difficult.

A naive estimate of the position of the edges (or a corresponding
fitting to the box-shape) could only be reliably performed if the
feature is significantly distinct. Including all backgrounds, this
is not the case. While the position of $E_{\rm max}$ is rather well
visible, reading off $E_{\rm min}$ is questionable since the low
edge is essentially smeared out by the SUSY background. The main
culprit is the irreducible SUSY background processes, where one of
the observed leptons ($e$ or $\mu$) comes directly from chargino
decays into $\ell\wt{\nu}_\ell$ and the other from the $\tau$ that
is the decay product of the chargino or even the stau, i.e.\
$\tilde{\chi}^+_1\to \tau^+\tilde{\nu}_\tau$ or $\tilde{\chi}^+_1\to
\tilde{\tau}^+ \nu_\tau \to \tau^+\nu_\tau\tilde{\chi}^0_1$ (see the
discussion in Sec.~\ref{sec:cutedge}). The energy spectrum of the
tau decay leptons is peaked at low energy. The transverse momentum
cuts, which are necessary to suppress the huge photon-induced SM
background, tweak these low-energy distributions in such a way
that the $E_{\rm min}$ edge becomes completely swamped.

If the chargino mass, however, was determined elsewhere (from e.g.~the
threshold scan in a clearly visible and distinguishable channel), the
$E_{\rm max}$ can be used to derive the sneutrino mass with the help of
\begin{equation}
m^{2}_{\wt{\nu}_\ell} = m_{\wt{\chi}^\pm_1}^{2} \lb
1-\frac{4\,E_\text{max}}{\sqrt{s}}\,\frac{1}{1+\beta} \rb \;.
\end{equation}
Taking the chargino mass as an input with a conservative estimate on
its error, $m_{\wt{\chi}^\pm_1}\,=\,184\,\pm\,1\,\GeV$,  as well as a
rough estimate for the upper lepton edge, $E_\text{max} = 24 \pm 2$
GeV, we obtain
\begin{\eqn}\label{eq:numassthr}
m_{\tilde{\nu}}\,=\,173.1\,\pm\,1.3\,\GeV,
\end{\eqn}
which agrees within large error with the input value of the
sneutrino mass $m_{\wt{\nu}} = 172.52$ GeV. The large error is the
direct consequence of the large error for $E_{\rm max}$.

For the purpose of the above analysis the edge $E_\text{max}$ and
its error has been determined ``by eye''~\cite{Lester:2005je}. It is
common when extracting the observed edges from plots, such as those
above, to fit a function to the endpoint in order to determine both
the precision and the accurate position.  However, when taking into
account ISR, beamstrahlung, background processes, off-shell and
interference effects and their response to kinematic cuts, the form
of fitting function is impossible to determine analytically. As
stressed in Ref.~\cite{Gjelsten:2004ki}, even in favorable cases
applying analytic functions too readily may lead to inaccurate measurements and
underestimated errors, since  endpoints can often exhibit tails or
get smeared. Therefore, in the next subsection we develop
a different experimental method of determining $E_{\rm max}$ that
allows for a more precise measurement of $m_{\wt{\nu}}$.


\subsection{Sneutrino mass determination from $\chi^{2}$ bin by bin
fitting}
\label{sec:chi2}

In order to use maximum information from energy distributions of the
final state leptons, we apply another method of determining the
sneutrino mass at $\sqrt{s} = 500$~GeV, where we perform a binned
$\chi^2$ fit. This is accomplished by using the MC true sample for
SPS1a' as the ``experimental data'' and generating MC control samples
for varying sneutrino masses within the range $171 \leq
m_{\tilde{\nu}} \leq 174$~GeV in a small window around the true
value. For this, we assume knowledge of the approximate sneutrino mass
range from one of the methods presented in the previous two sections.
Despite the fact, that this method can also be used  in the case
of non-degeneracy in the first two generation sneutrinos, we keep that
assumption for the following analysis. In a non-degenerate case, the
fitting had to be applied for each  lepton energy
distribution separately. Here, however, we combine the information from both
histogram fits. The tau sneutrino mass is left at its nominal SPS1a'
value. For the histograms, we use a conservative value for the binning
of 1~GeV. The signal and all SUSY backgrounds are considered for the
analysis.

\FIGURE{
    \includegraphics[width=.42\textwidth]{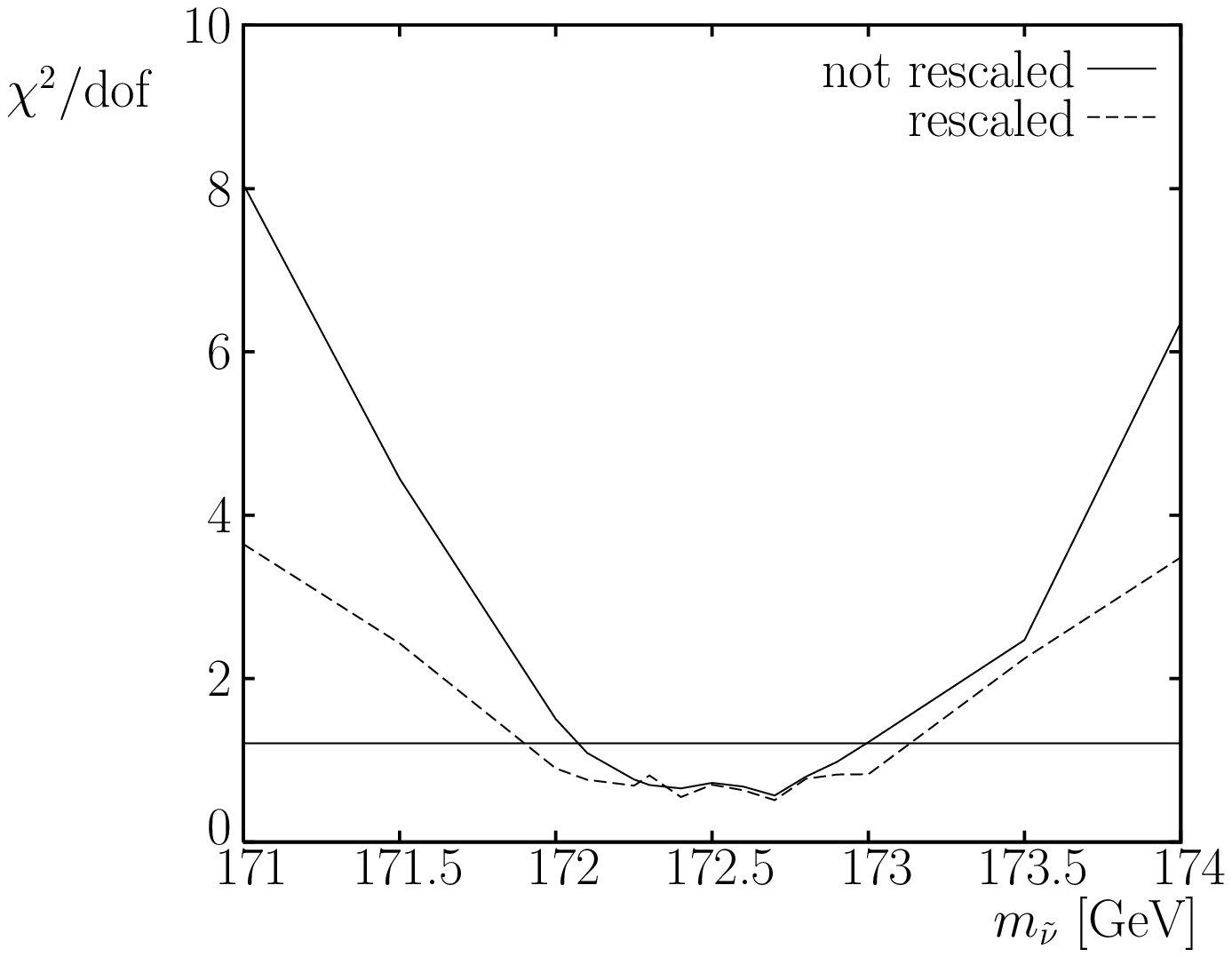}
    \qquad
    \includegraphics[width=.42\textwidth]{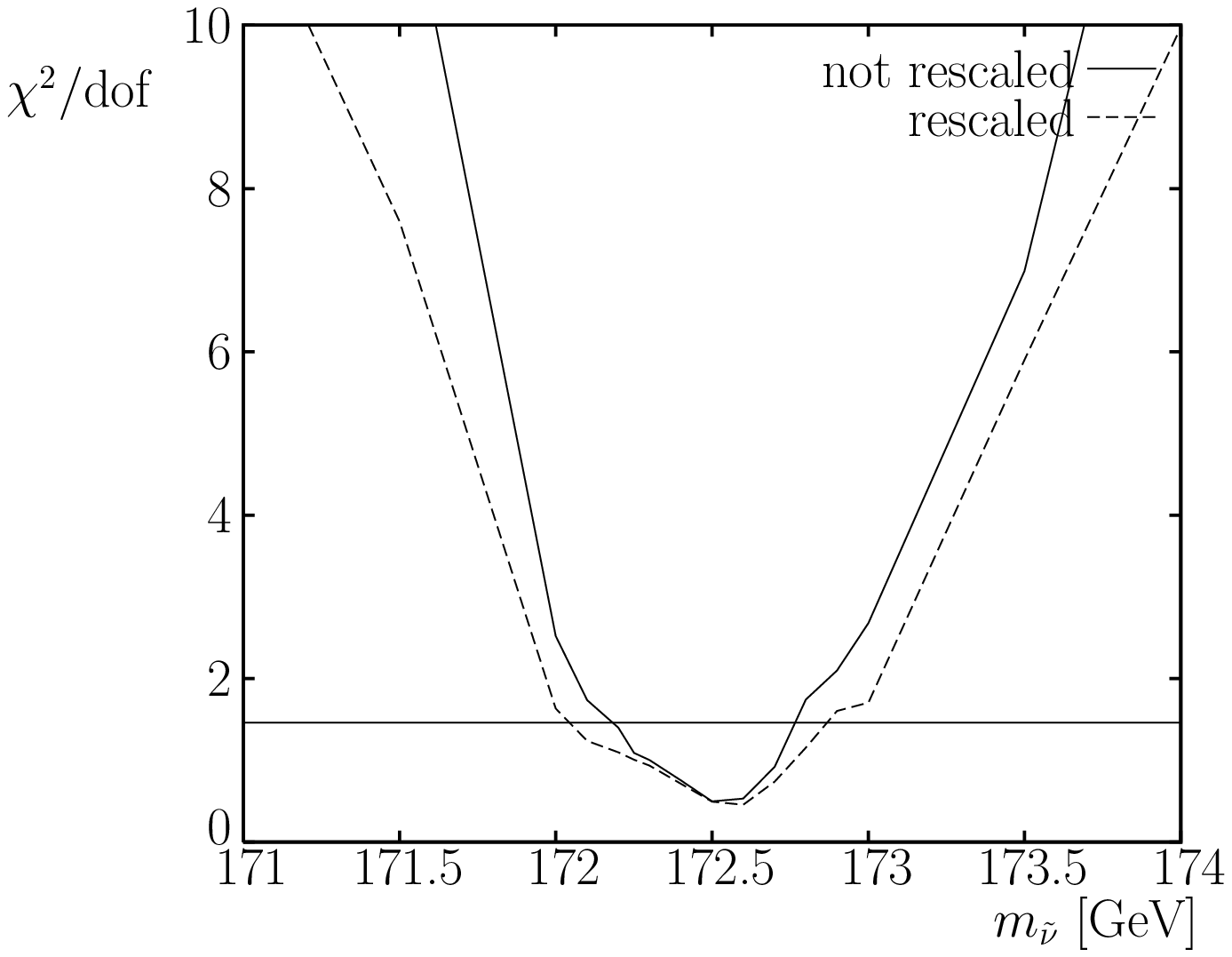}
\caption{\label{fig:chi2} $\chi^2$/dof distributions
  for the sneutrino mass determination. Left: fitting the
  whole energy spectra of both $e$ and $\mu$, right: fitting the bins
  around the upper edges. Dashed line are with, solid lines without rescaling
  (see text for explanations). Horizontal lines denote a significance
  level of $10\%$. }
}

Mathematically, $\chi^2$ is the best estimator whenever the
uncertainties of each bin are Gaussian. Since this is the case
(Poisson statistics from Monte Carlo with large enough event numbers),
we use a $\chi^2$ fit for the mass determination. We are aware of the
fact that matrix element methods give the best statistical means for
a particle mass determination (as e.g.~for the top mass determination
at Tevatron and LHC). However, such an analysis is beyond the scope of
the present paper. The $\chi^2$ statistics used in our fit has the
following form
\begin{equation}
\label{eq:chisq}
\chi^2(m_{\wt\nu}) =
\sum_i \frac{
        (N_i^{\mathrm{tr}}    - N_i^{\mathrm{con}} (m_{\wt\nu}) )^2}
        {N_i^{\mathrm{tr}} + N_i^{\mathrm{con}}(m_{\wt\nu})} \;,
\end{equation}
where the sum runs over bins in energy distributions of electron and
muon, $N_i^\mathrm{con}(m_{\tilde{\nu}})$ is the number of expected events in the
$i$-th bin from our control sample as a function of the sneutrino mass, and
$N_i^\mathrm{tr}$ is the number of
observed events in the $i$-th bin for the true SPS1a' sneutrino mass;
 statistical errors are added in quadrature.
Using Eq.~(\ref{eq:chisq}), we calculate $\chi^2$ statistics for
each of the control sample histograms.

We perform two types of fits. In the first, we use all bins in the
full energy range $2-40$~GeV, yielding 76 degrees of freedom (dof)
when combining electron and muon data. In the second, we use only
bins in the vicinity of the upper edge, which is assumed to be
experimentally determined -- at least approximately -- by e.g.~a
total cross section measurement (see Sec.~\ref{sec:masscross}). As a
window, we use $\pm 4$~GeV around the expected upper energy edge of
the decay leptons. The second method amounts to 18 dof for the
$\chi^2$ statistics. We perform each of the two fits in two
variants: using the absolute numbers of events per bin in the
control sample, and in a rescaled version. In the latter  we
normalize the control sample to the number of true events, i.e.\ we
define for each bin
\begin{equation}
N_i^\mathrm{con,rescaled} = N_i^\mathrm{con}
\frac{N_\mathrm{tot}^\mathrm{tr}}{N_\mathrm{tot}^\mathrm{con}}\;,
\end{equation}
where $N_\mathrm{tot}^\mathrm{tr}$ and $N_\mathrm{tot}^\mathrm{con}$
are the total number of events in the true and control samples,
respectively. Since the normalization to the true sample imposes an
additional constraint, it decreases effectively the number of dof by
one. In this case the $\chi^2$ is sensitive only to the shape of
energy distributions  which does not rely much on the knowledge of
the rest of the SUSY spectrum. Therefore, the error for the
sneutrino mass determination is dominated by statistics, and does
not suffer too much from parametric uncertainties.

\FIGURE{
    \includegraphics[width=.42\textwidth]{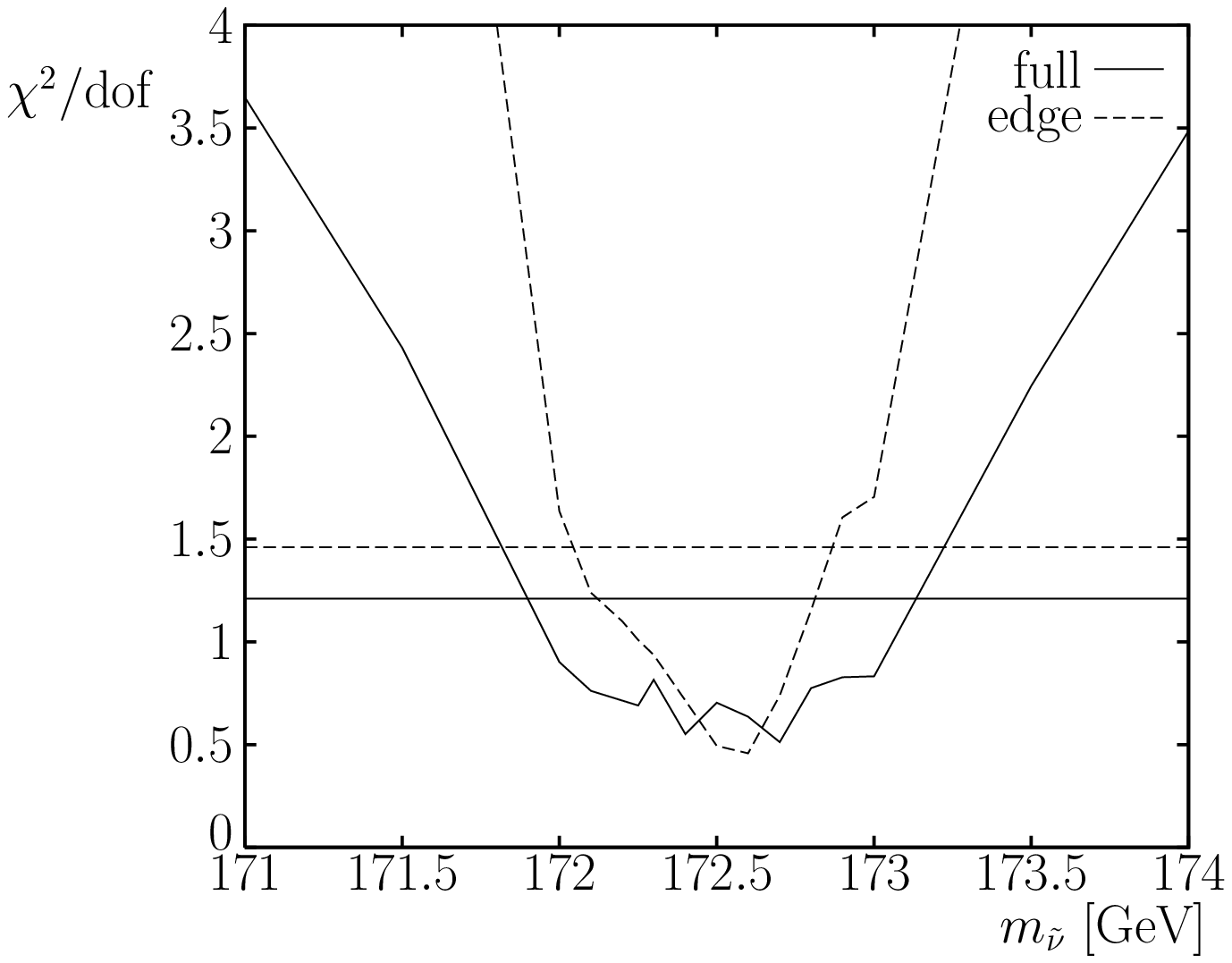}\qquad
    \includegraphics[width=.42\textwidth]{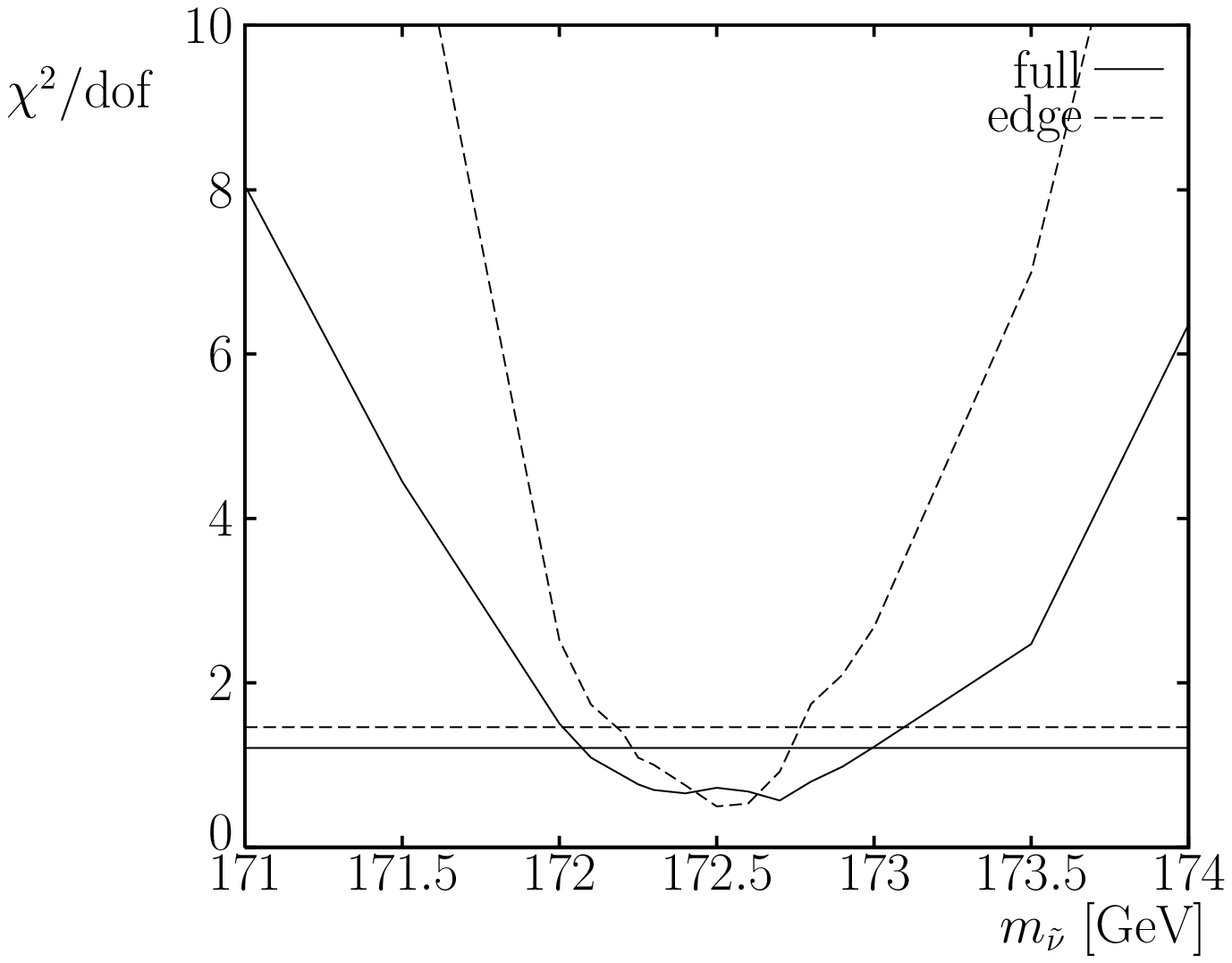}
\caption{\label{fig:chi2_resc} $\chi^2$/dof distributions for the
sneutrino mass determination. Fits for the complete energy spectra
(solid) and in the window around the $e$ and $\mu$ edges (dashed
lines). Left panel for rescaled number of events in the bins, right
panel without rescaling (see text for explanations). Horizontal
lines denote significance level of $10\%$ for 76 (solid, full range)
and 18 (dashed, restricted window) dof, respectively. }
}

The plots in Fig.~\ref{fig:chi2} for the $\chi^2$/dof  calculated
according to Eq.~(\ref{eq:chisq}) compare the two methods with and
without rescaling the number of events in each bin. On the left
hand side, the fits over the whole energy range are shown, whereas the
right plot shows the result of restricting the fit to the window
around the edge. The horizontal lines denote a $10\%$ significance
level for the $\chi^2$/dof. In Fig.~\ref{fig:chi2_resc} we illustrate
the difference between $\chi^2$/dof in the full energy range versus the restricted energy
window fits with (left panel) and  without (right panel) rescaling.

As can be seen from both Figures~\ref{fig:chi2} and
\ref{fig:chi2_resc}, the $\chi^2$ distributions show significant
statistical fluctuations, especially around their minima. These
arise due to statistical fluctuations in the generation of the MC
control samples, as well as those in the true MC sample, where the
latter would correspond to the true experimental statistical
uncertainty.

In order to assess the possible accuracy in the sneutrino mass
determination, we test the hypothesis whether the control sample for
a given $m_{\wt{\nu}}$ coincides with the MC true (``experimental'')
data.
We define the compatibility region for the sneutrino mass as
the range where the null hypothesis cannot be rejected at the
significance level of $10\%$. Note, that this approach does not
directly translate to a confidence level (CL) interval for the
sneutrino mass, but rather gives a conservative estimate. Since we
did not include any systematic and parametric uncertainties, we
consider such a conservative estimate a reasonable alternative to
the usual procedure of taking the respective region around the
minimum of the $\chi^2$ distribution.

\TABLE
{
\begin{tabular}{|c||c|c|c|c|}\hline
 method $\rightarrow$ & full range & full range & edge & edge\\
 \cline{1-1}
 limit $\downarrow$ & rescaled & not rescaled & rescaled & not
rescaled\\
 \hline \hline
 lower & 171.9 & 172.1  & 172.1  & 172.2 \\ \hline
 upper & 173.1 & 173.0  & 172.9  & 172.8 \\ \hline
\end{tabular}
\caption{Upper and lower limits of the compatibility range for the
sneutrino mass (in GeV) corresponding to the significance level of
$10\%$ of testing the null hypothesis using different methods for
$\chi^2$ fits. The true value for the sneutrino mass is
$m_{\wt{\nu}} = 172.52$~GeV. \label{tab:chisq_results}}
}

The results of $\chi^2$ fits using different approaches discussed
above are collected in Table~\ref{tab:chisq_results}. In general,
the resulting accuracy is of the order of $0.5$~GeV. The best
precision is obtained for edge fitting without rescaling samples. Edge
fitting quite generically gives better results. Although,
when using the full range of all 76 bins in the fit, we utilize more
information about the distribution, a lot of it is irrelevant for
sneutrino mass determination, hence giving a somewhat broader minimum
of the $\chi^2$ function. As mentioned before, since we loose some
information concerning the total cross section (or, on the
contrary, dependence on the SUSY model) when rescaling data, therefore
-- as expected -- in this case the fits give lower precision.
On the other hand, rescaling minimizes effects which can result from
parametric uncertainties on the SUSY model.

\vspace{3mm}

In summary, we studied three different methods of sneutrino mass
determination, including all SM and SUSY backgrounds. A first
estimate, assuming the rest of the spectrum is known, can be
obtained from the total number of SUSY events (signal and all
backgrounds) after the application of cuts.

Alternatively, we can use purely kinematic information from the two
subsequent two-body decays of the chargino and the sneutrino; in
that case, the error is twice as large as the one obtained
from the total cross section, but does not depend as strongly on the
specific SUSY model. The main difficulty of this approach is
defining the position of the upper edge, which is smeared by cut
effects, ISR and beamstrahlung.

The most sophisticated method that we developed employs a fit of the
``experimental'' lepton energy spectra by template distributions
generated with varying sneutrino mass. The method allows one to include
or exclude additional knowledge on the SUSY model. Here, the precision
of the sneutrino mass determination can be reduced to $\pm
0.4\; (\pm 0.3)\,\GeV$, when performing edge fits without (with)
additional information on the SUSY spectrum, respectively. We used
the simplest version of this method; a full analysis would require
matrix element reweighting for a correct treatment of the
statistical properties of ``true data'' and ``control samples'' with
varied sneutrino masses. On top of that, a treatment of systematic
uncertainties (like detector effects) is mandatory to get a decisive
answer on the final precision of the sneutrino mass determination.
This should be a part of an accompanying experimental study.


\section{Conclusions and Outlook}
\label{sec:outlook}

In this paper we have investigated the problem of the discovery of the invisibly 
decaying sneutrino and its mass determination in a realistic ILC
environment. We considered the SPS1a' scenario in which sneutrinos
are lighter than the lightest chargino and next-to-lightest
neutralino. Since decay modes with charged particles in the final
state are of higher order, and thus strongly suppressed, sneutrinos
decay invisibly via a tree-level process into the lightest
neutralino and neutrino. As a result, sneutrino masses cannot be
measured by a threshold scan. Therefore we have analyzed the
opportunity to measure sneutrino masses from the kinematics of the
chargino pair production at the ILC followed by two-body chargino
decays $\tilde{\chi}^\pm_1\to \ell^\pm\tilde{\nu}_\ell^{(*)}$ (with
$\ell=e,\,\mu$).

The two-body chargino decays generate sharp edges in the lepton
energy spectra which depend on chargino and sneutrino masses. The
edges, however, get smeared by ISR and beamstrahlung. Moreover, the
signal process is overwhelmed by a plethora of background processes
generated by both SUSY and SM mechanisms. Since the  selection
procedure required to enhance the signal/background ratio distorts
the lepton spectra from the signal and from various background
processes in different manners, particular attention has been paid
to the pollution coming from these effects.

We have considered all SUSY and SM background processes that could
be of relevance to the analysis. The main observation is that the
leptons coming from the leptonic decays of the final $\tau$'s
(copiously produced by SUSY and SM processes) practically wash out
the lepton low-energy edge preventing precise determination of both
the chargino and sneutrino masses. If the chargino mass, however, is
predetermined from a different observable, the high-energy edge
alone is sufficient to measure the sneutrino mass with an accuracy
of order a GeV. Using a $\chi^2$ fit of  control samples
generated for varying sneutrino mass to the ``experimental'' one at
the nominal value we find that the error can be further reduced by a
factor~4.

The calculation has been performed at tree-level, demonstrating the
capability of \whizard\ to generate exclusive final states using
full matrix elements for production and decay, including off-shell
and interference effects. The expected high experimental precision
at the ILC, however, calls for more precise theoretical methods.
Recently the chargino pair production process at one-loop has been
incorporated into \whizard\ \cite{chargino_nlo}. One-loop
calculations of the chargino decay process are also
available~\cite{decayNLO}, so a full NLO calculation of the signal
process, which has to be incorporated in the simulation, is a
possible future improvement of the present study.

\bigskip


\subsection*{Acknowledgements}

The authors are grateful to Philip Bechtle, Seong Youl Choi, Klaus
Desch, Ayres Freitas, Hans-Ulrich Martyn, and Peter Zerwas for
fruitful discussions. Special thanks go to Malgorzata Worek and
Zbigniew Was for their help concerning tau decays. JR thanks the
Aspen Center of Physics for their hospitality, and JK and KR
acknowledge warm hospitality extended to them at the NORDITA program
``TeV scale physics and dark matter''.

JR was partially supported by the Bundesministerium f\"ur Bildung
und Forschung, Germany, under Grant No. 05HA6VFB. TR was supported
by the DFG SFB/TR9 ``Computational Particle Physics''.  JK and KR
were supported by the Polish Ministry of Science and Higher
Education Grant No~1~P03B~108~30. This research was supported by the
Helmholtz-Alliance ``Physics at the Terascale'',  the EU Network
MRTN-CT-2006-035505 ``Tools and Precision Calculations for Physics
Discoveries at Colliders", the EC Programme MTKD--CT--2005--029466
``Particle Physics and Cosmology: the Interface'', as well as by the
Helmholtz-Gemeinschaft under Grant No. VH-NG-005 at an early stage of
the work. 

\bigskip\bigskip



\begin{appendix}
\numberwithin{figure}{section} \numberwithin{table}{section}

\section{The SPS1a' parameter point}
\label{sec:sps1ap}

In this appendix we collect the features of the SPS1a' parameter
point which are relevant for our analysis. The spectrum of
supersymmetric particles is given in Fig.~\ref{fig:sps1ap}. The
masses and decay branching fractions, generated using
\spheno~\cite{spheno} and \sdecay~\cite{sdecay}, are given in
Tab.~\ref{tab:sps1ap}. The consistency with low-energy constraints
has been checked with \spheno~\cite{spheno} and the cold dark matter
constraint with \micromegas~\cite{micromegas}.

\TABLE[ht]
{
\renewcommand{\arraystretch}{1.}
  \begin{tabular}{|c|c||c|c||c|c|}\hline
    \text{Parameter}  & \text{Value} & \text{Parameter} & \text{Value} &
    \text{Decay} & \text{BR} \\\hline\hline
    & & $m_h$             & 116.0 GeV  &
    $\wt{\chi}_1^+ \to \wt{\tau}_1^+ \nu_\tau$    &   53.6 \%
    \\
    & & $m_A$             & 424.9 GeV  &
    $\wt{\chi}_1^+ \to \wt{\nu}_\ell \ell^+$          &   13.3 \%
    \\
    $m_0$             &   70 GeV     & $m_{\wt{\chi}_1^0}$    &  97.7 GeV  &
    $\wt{\chi}_1^+ \to \wt{\nu}_\tau \tau^+$    &   18.5 \%
    \\
    $m_{1/2}$         &  250 GeV  & $m_{\wt{\chi}_2^0}$    & 183.9 GeV  &
    $\wt{\chi}_1^+ \to \wt{\chi}^0_1 W^+$             &    1.3
    \%  \\\cline{5-6}
    $\tan\beta$      &   10    & $m_{\wt{\chi}_1^\pm}$  & 183.7 GeV &
    $\wt{\ell}^-_R \to \wt{\chi}_1^0 \ell^-$             &  100.0
    \% \\\cline{5-6}
    $\sgn\mu$         &   +       & $m_{\wt{\chi}_2^\pm}$  & 415.4 GeV &
    $\wt{\ell}^-_L \to \wt{\chi}_1^0 \ell^-$             &   92.5
    \% \\\cline{3-4}
    $A_0$   & $-300$ GeV & $m_{\wt{\ell}_R}$ & 125.3 GeV &
    $\wt{\ell}^-_L \to \wt{\chi}_1^- \nu_\ell$        &    4.9
    \%
    \\
    & & $m_{\wt{\ell}_L}$ & 189.9 GeV &
    $\wt{\ell}^-_L \to \wt{\chi}_2^0 \ell^-$    &    2.6 \%
    \\\cline{1-2}\cline{5-6}
    & & $m_{\wt{\nu}_\ell}$ & 172.5 GeV &
    $\wt{\chi}_2^0 \to \wt{\tau}_1^\pm \tau^\mp$   &   57.8 \%
    \\\cline{3-4}
    BR($b\to s\gamma$) & $4.69\cdot 10^{-4}$ &  $m_{\wt{\tau}_1}$ & 107.9 GeV &
    $\wt{\chi}_2^0 \to \wt{\nu}_\tau \bar{\nu}_\tau$ + cc.   &   15.2
    \%
    \\
    $(g-2)_\mu$    & $8.041\cdot 10^{-9}$  & $m_{\wt{\tau}_2}$ & 194.9 GeV &
    $\wt{\chi}_2^0 \to \wt{\nu}_\ell \bar{\nu}_\ell$   + cc.      &   11.1
    \%  \\
    $\Delta\rho$     & $1.87 \cdot 10^{-4}$ & $m_{\wt{\nu}_\tau}$ & 170.5 GeV &
    $\wt{\chi}_2^0 \to \wt{\ell}^\pm \ell^\mp$        &    2.4
    \% \\\cline{3-6}
    $\Omega h^2$    &  0.10 & $\Gamma(\wt{\chi}^\pm_1) $ & 77.3 MeV &
    $\wt{\tau}^-_2 \to \wt{\chi}^0_1 \tau^-$ & 86.8\% \\ 
    & & $\Gamma(\wt{\nu}_\ell) $ & 121.5 MeV & $\wt{\tau}^-_2 \to
    \wt{\chi}^0_2 \tau^-$ & 4.6\% \\ 
    & & $\Gamma(\wt{\nu}_\tau) $ & 117.4 MeV & $\wt{\tau}^-_2 \to
    \wt{\chi}^-_1 \nu_\tau$ & 8.6\% \\ 
    \hline
    \hline
  \end{tabular}
  \caption{\label{tab:sps1ap} Details of the SPS1a' parameter point. $\ell$ stands for $e$ or $\mu$.}
}

\FIGURE[ht]{
\includegraphics[height=7.5cm]{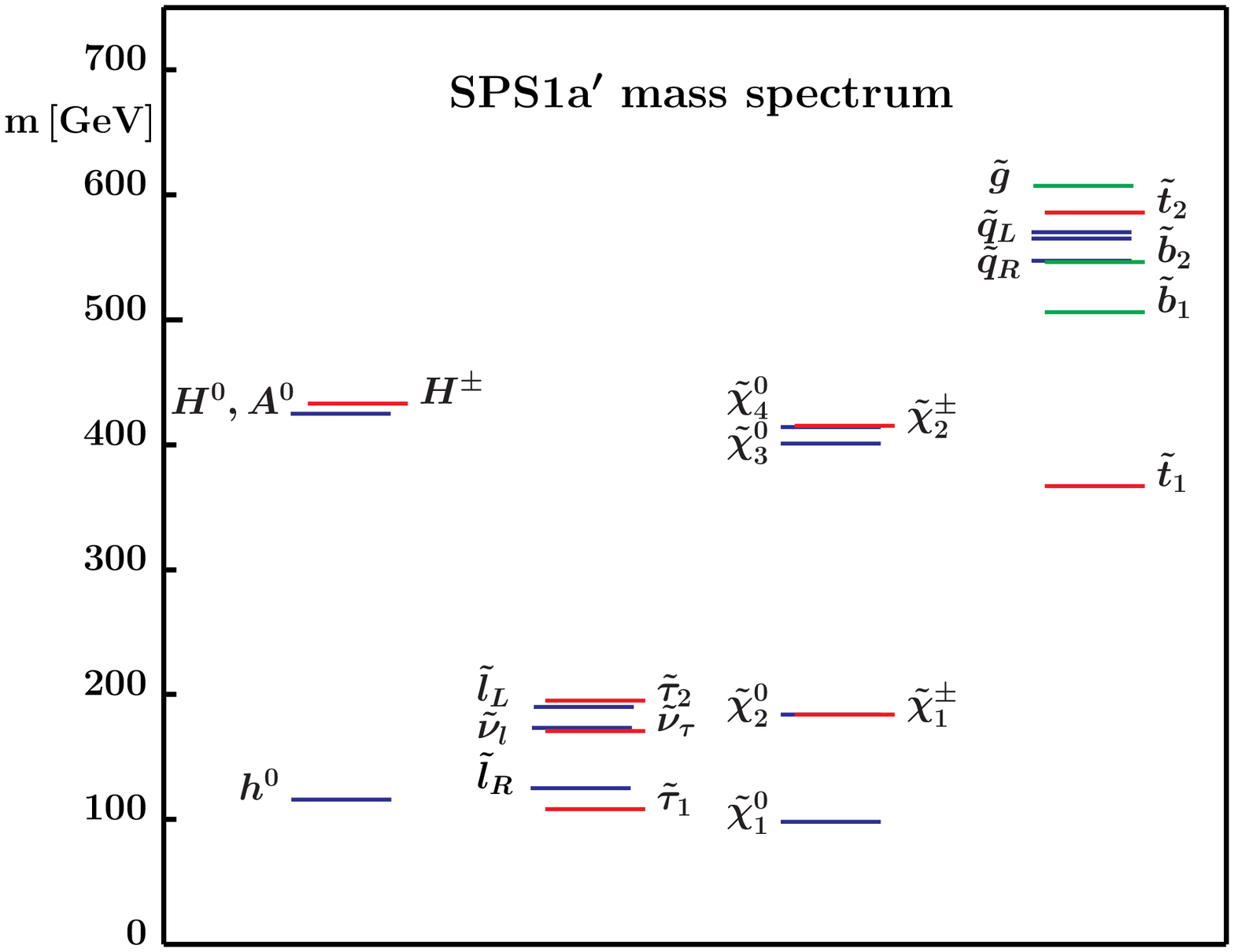}
\caption{\label{fig:sps1ap} Particle spectrum of the SPS1a'
parameter point.}
}

\section{Comparison of EPA with the exact matrix element calculation }
\label{sec:epa}

Photon-induced processes at linear colliders are commonly simulated
using the equivalent photon approximation (EPA)
\cite{Budnev:1974de}. The \whizard\ code allows one to use the
exactly generated matrix element (cf.\ Sec.~\ref{sec:simu}) and
therefore provides means to test the quality of EPA. In the
following we will expose the main differences between EPA and the
exact matrix element calculations (EME). As an example we take the
photon-induced tau-pair production process with successive leptonic
tau  decays.

\FIGURE{
    \includegraphics[width=.37\textwidth]{compexepa.1}\qquad\qquad
    \includegraphics[width=.37\textwidth]{compexepa.3}
\vspace{2mm}
\caption{\label{fig:compexepadist} Lepton energy distributions from
  the photon-induced $\tau$ pair production processes (left: electron, right:
  muon) using the equivalent photon approximation (solid, black) and the
  exact matrix element (dotted, red). }
}

\FIGURE{
\includegraphics[width=.37\textwidth]{compexepa.2}\qquad\qquad
    \includegraphics[width=.37\textwidth]{compexepa.4}
\vspace{2mm} \caption{\label{fig:compexepadiff} As
  Fig.~\ref{fig:compexepadist}, but now plotting the difference
  between the EME and the rescaled EPA, $N^{\rm EME}_i-N_i^{\rm
  res}$. The 1-$\sigma$ error from the exact result is also given
  (dotted, red).  } 
}

\TABLE{
\begin{tabular}{|c||r|r|r|r|r||r|}
\hline
Process &
presel\phantom{no} & no $p_{\perp}$\phant & no $E$\phant &
no $\theta$\phant & no $\Delta\phi$\phantom{no} &
all cuts
\\ \hline\hline
\text{exact}&787927&22685&10&14&867&10\\
\text{EPA}&767322&27974&--&--&1546&--\\
\hline
\end{tabular}
\caption{\label{tab:cuteffsepa} Comparison of cut efficiencies as in 
  Table \ref{tab:cuteffs} for both EPA and exactly generated
  photon-induced $\tau$'s, normalized to 1 million events before
  preselection cuts.}
}

The total cross section for photon-induced taus with their successive
decay after preselection at 500 \GeV are as follows: 
\begin{\eqn}
\sigma_{\gamma\tau}^{\text{EPA}}\,=\,25.495(4)\,\pb,\qquad
\sigma_{\gamma\tau}^{\text{EME}}\,=\,21.392(70)\,\pb, 
\end{\eqn}
so the EPA overestimates the number of generated taus.
Figure~\ref{fig:compexepadist}  shows the difference between lepton
energy spectra generated with EPA and EME for electrons (left) and
muons (right). The shapes of the distributions are roughly the same,
so one may be tempted to account for the larger EPA  distribution by
a simple rescaling factor $K=\sigma^{\rm EME}/\sigma^{\rm EPA}$,
i.e. taking $N_i^{\rm res}= K\, N_i^{\rm EPA}$ in the $i$-th bin.
Figure~\ref{fig:compexepadiff} shows the difference between EME and
such rescaled EPA distributions demonstrating  that the  rescaled EPA
clearly underestimates the number of low energy leptons. For
investigations that rely on the lepton spectra at low energies  this
feature becomes crucial. In our analysis, some of the cuts were
tailored to remove the low-energy leptons from photon-induced
background, in particular the $p_\perp$ and $\Delta\phi$ cuts.
Table~\ref{tab:cuteffsepa} shows the response of a sample of 10$^6$
EPA and EME events each to the cuts. Employing the EPA instead of
EME calculations would lead to an artificial reduction of the SM
background by roughly 13\%; for an integrated luminosity of 1
ab$^{-1}$ this corresponds to a 5$\sigma$ deviation from the
expected SM background. It is therefore crucial to use the exact matrix elements in
calculations of photon-induced processes.



\end{appendix}


\baselineskip15pt


\end{document}